\documentclass[aps,prd,10pt,nofootinbib,twocolumn,eqsecnum,showpacs,showkeys,superscriptaddress,preprintnumbers,altaffilletter]{revtex4-2}
\usepackage{enumitem} 
\usepackage{graphicx}
\usepackage{amsmath} 
\usepackage{multirow}
\usepackage{hyperref}
\usepackage{soul}
\usepackage{graphicx}
\usepackage{dcolumn}
\usepackage{amssymb}
\usepackage{amsfonts}
\usepackage{amsbsy}
\usepackage{color}
\usepackage{rotating}
\usepackage[english]{babel}
\usepackage{multirow}
\usepackage{float}
\usepackage{cleveref}
\usepackage{diagbox}
\usepackage{subcaption}
\usepackage[thinc]{esdiff}

\usepackage{amsmath,latexsym,mathtools}
\usepackage{standalone}
\usepackage{import}
\usepackage{tikz} 
\tikzset{
  pics/carc/.style args={#1:#2:#3}{
    code={
      \draw[pic actions] (#1:#3) arc(#1:#2:#3);
    }
  }
}

\usepackage{booktabs}

\newcolumntype{x}[1]{>{\centering\arraybackslash\hspace{0pt}}p{#1}}

\newcommand{\be}{\begin{equation}}
\newcommand{\ee}{\end{equation}}
\newcommand{\bea}{\begin{eqnarray}}
\newcommand{\eea}{\end{eqnarray}}
\newcommand{\beal}{\begin{aligned}}
\newcommand{\eeal}{\end{aligned}}
\newcommand{\bi}{\begin{itemize}}
\newcommand{\ei}{\end{itemize}}

\newcommand{\tb}{ \textcolor{blue}}

\hypersetup{
    colorlinks=true,
    linkcolor=blue,
    filecolor=magenta,
    urlcolor=cyan,
    citecolor=cyan
}

\begin{document}

\title{Finite time pseudo-rip singularity in cosmology}

\author{Mariusz P. D\c{a}browski}
\email{mariusz.dabrowski@usz.edu.pl}
\affiliation{Institute of Physics, University of Szczecin, Wielkopolska 15, 70-451 Szczecin, Poland.}
\affiliation{National Centre for Nuclear Research, Andrzeja So{\l}tana 7, 05-400 Otwock, Poland.}
\affiliation{Copernicus Center for Interdisciplinary Studies, Szczepa\'nska 1/5, 31-011 Krak\'ow, Poland.}
\author{Teodor Borislavov Vasilev}
\email{teodorbo@ucm.es}
\affiliation{Departamento de F\'isica Te\'orica and IPARCOS, Universidad Complutense de Madrid, E-28040 Madrid, Spain.}

\date{\today}

\begin{abstract}
By studying new decelerating sudden future singularity (SFS) universe, we report finding a novel type of cosmological singularity which we dub a finite time pseudo-rip (FTPR). 
In contrast to the new SFS model, where the expansion is decelerating before reaching the pressure singularity, the FTPR scenario is preceded by a super-accelerated phantom phase. Our claim is based on the thorough study of the energy conditions showing the violations of all of them for a FTPR, and only the dominant energy one for an SFS. Application of the so-called Raychaudhuri averaging shows that, alike within the requirement of geodesic completeness, these singularities are weak in the sense of this definition. We study the properties of the models including the behaviour of the cosmological horizons presented in the appropriate Penrose diagrams. Finally, we introduce a FTPR model containing standard radiation and dust fluids that can mimic the past expansion history of $\Lambda$CDM though facing a future pressure singularity.

\end{abstract}

\maketitle

\section{Introduction} 

Since the discovery of the accelerated expansion of the Universe and the postulation of dark energy \cite{Riess98, Perlmutter99}, together with recent reports of observational tensions \cite{tension}, a variety of models have been proposed as alternatives to the standard $\Lambda$CDM cosmology. These models aim to provide a viable late-time behaviour while alleviating some of the conceptual problems of the cosmological constant \cite{CC,CC2}. However, deviations from the standard model are usually accompanied with theoretical and observational challenges of their own \cite{+LCDM,+LCDM2}. In particular, the exploration of strongly negative pressure components, such as phantom energy \cite{Caldwell2002,Caldwell2003,MPD2003}, has revealed the possibility of future singularities like the big-rip (BR), leading to the destruction of all bound structures, and has motivated numerous extensions that fully or partially violate the Hawking--Penrose energy conditions (EC) in General Relativity \cite{HE}. These singularities showed up consecutively following Barrow's paper on the sudden future singularity (SFS) models \cite{Barrow2004,Barrow2004a} and later have been developed by other researches. One of the first systematic attempt to classify all possibilities was given in Ref. \cite{Nojiri2005}, called types I to IV, and later expanded, among others, in Ref. \cite{MPD2014CC}. In general, models which fall outside the classical big-bang (BB) singularity are also known under the name of ``exotic'' singularities due to their diverse behaviour in comparison to their original canon. 

In this paper, we investigate two new exotic singularity models by postulating some specific behaviour of the Hubble parameter related to the cosmological fluid energy density: one violating the dominant energy condition only and another violating all the energy conditions. Both models lead to a future curvature singularity in a finite value of cosmological time. These singularities may be classified as an SFS and a new type of event, hereafter referred to as a finite-time pseudo-rip (FTPR), which--up to certain subtleties discussed in the paper--differs from other intrinsic phantom scenarios such as the pseudo-rip (PR) \cite{PR,PR1}, the little sibling of the big-rip (LSBR) \cite{LSBR}, the little-rip (LR) \cite{LR}, the BR \cite{Caldwell2002,Caldwell2003,MPD2003} or the big-freeze (BF)  \cite{BF,BF1}.

Our paper is organized as follows. In Section \ref{models} we present our universe models which admit the exotic singularities. In Section \ref{ECond} we explore these models in terms of the energy conditions. In Section \ref{Penrose} we present the Penrose diagrams for the models which allow to give insight into their asymptotic structure. In Section \ref{Raych} we test the models using the Raychaudhuri weak singularity criterion. In Section \ref{sec:LCDM FTPR}, we present an exotic singularity model that fully mimics the past background expansion of the standard cosmological $\Lambda$CDM scenario, while evolving towards a future FTPR pressure singularity. In the last Section \ref{Concl} we present our conclusions. 


\section{New exotic singularity models}
\label{models}
	
\subsection{A decelerating sudden future singularity model}
	
We propose a new model of SFS \cite{Barrow2004} (or type II \cite{Nojiri2005,big-brake,Hendry2007,Keresztes2009,Keresztes2010}) universe which is described by the Hubble rate
\begin{eqnarray}
\label{eq:model}
H[a(t)] \equiv \frac{\dot{a}}{a} = H_\star \frac{a_s^n}{a^{ \frac{m}{2}}(t)} \left( 1 - \frac{a(t)}{a_s} \right)^n ,
\end{eqnarray}
where $a(t)$ is the scale factor, and $H_\star$ and $a_s$ are positive constants. We also consider $m$ and $n$ to be positive exponents. The former allows us to mimic the standard cosmological components in the limit $a\to0$, corresponding to $m=6$ (stiff fluid), $m=4$ (radiation), $m=3$ (dust), $m=2$ (cosmic strings or spatial curvature), and $m=0$ (cosmological constant). Thus, in order to avoid phantom behaviour out of this factor, we assume \bea m \geq 0.\eea 
For the latter exponent, we further assume the restriction
\be\label{eq:n_SFS}
0<n<\frac12,
\ee
which will be explained below.   

From this expression for the Hubble rate, the second derivative of the scale factor reads
\bea
\label{eq:ddot a}
&& \frac{\ddot{a}}{a} = - H_\star^2 a_s^{2n} a^{-m} \left( 1 - \frac{a}{a_s} \right)^{2n-1}  \nonumber \\
&& \times \left[ \left( \frac{m}{2} - 1 \right)  - \left( \frac{m}{2} -1 -n \right) \frac{a}{a_s} \right]. 
\eea
To discuss on the phenomenology of this model is useful to obtain the effective energy density and pressure induced on a flat spatial geometry isotropic  Friedmann-Lema\^itre-Robertson-Walker (FLRW) cosmology. These are
\bea
\varrho &=& \frac{3}{8\pi G} H^2 , \label{rho1} \\
p &=& - \frac{c^2}{8 \pi G} \left( 2 \dot{H} + 3 H^2 \right) \label{p1},
\eea
respectively, being $G$ the Newton's gravitational constant, $c$ the speed of light and $\dot H$ represent the cosmic time derivative of the Hubble rate. Moreover, using equations (\ref{eq:model})-(\ref{p1}), it is possible to produce the equation of state  (for a flat $k=0$ universe) as
\begin{align}
\label{eq:EoS} 
w(a(t)) \equiv \frac{p}{\varrho c^2} = \frac{ m-3 -( m-3-2n)\frac{a(t)}{a_s}}{3\left(1-\frac{a(t)}{a_s}\right)}  ,
\end{align}
which plays the role of a time-dependent barotropic index.

\begin{figure*}

\begin{subfigure}{\columnwidth}
\includegraphics[width=\textwidth]{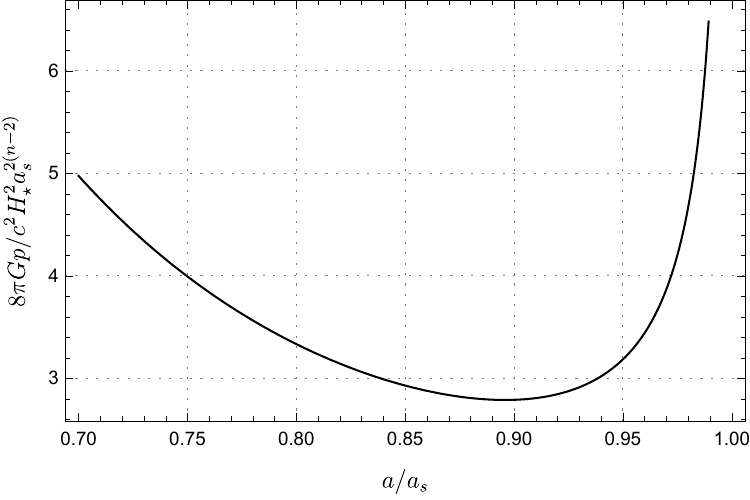}
\end{subfigure}
\hfill
\begin{subfigure}{\columnwidth}
\includegraphics[width=\textwidth]{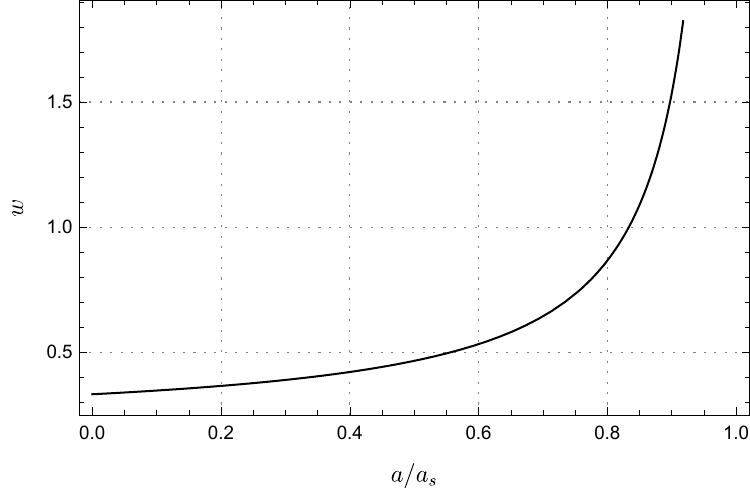}
\end{subfigure}
\caption{The evolution of the decelerating model (\ref{eq:model}) for $n=0.2$ and $m=4$ (radiation domination at $a\approx 0$). Left panel: the late-time evolution of the pressure $p$ given by (\ref{eq:EoS}).  Right panel: the evolution of the equation of state parameter $w$ according to expression (\ref{eq:EoS}). 
\label{fig:EoS}}
\end{figure*}

It is straightforward to check that the limit  $a \to 0$ corresponds to a BB initial singularity, at which $\varrho$ and $p$ diverge, whereas the corresponding equation of state parameter behaves as $w\approx(m-3)/3$  (see Fig. \ref{fig:EoS} for $m=4$). This is a new feature of our parametrisation that is not present in some previously addressed SFS models; see, for instance, Ref. \cite{GRG2018} where spatial-curvature-like effects ($m=2$) dominated at early stages in the expansion. For the limit $a\to a_s$, on the other hand, the Hubble rate (\ref{eq:model}) vanishes if $n>0$, says constant for $n=0$ or diverge for $n<0$.
The acceleration of the scale factor (\ref{eq:ddot a}) vanishes for $n>1/2$, converges to a constant for $n=1/2$ or diverges to infinity for $n<1/2$. Thus, for $n\in(0,1/2)$ the effective energy density vanishes whereas the effective pressure diverges at $a_s$. As advanced in expression (\ref{eq:n_SFS}), this behaviour correspond to a SFS or Type II singularity; see, for instance, \cite{Nojiri2005}. Interestingly, the pressure starts from infinity at BB, then diminishes reaching minimum in order to blow up to (positive) infinity again at (at least) the SFS (cf. Fig. \ref{fig:EoS}). This signals a decelerating behaviour near the SFS. Nevertheless, the equation (\ref{eq:ddot a}) confirms that the model at hands represents, in fact, an ever decelerating universe since $\ddot a$ is always negative for $a\in(0,a_s)$. In our approach, SFS plays the role of an extra fluid in the universe apart from the radiation.

\begin{figure}
\centering
\includegraphics[width=\columnwidth]{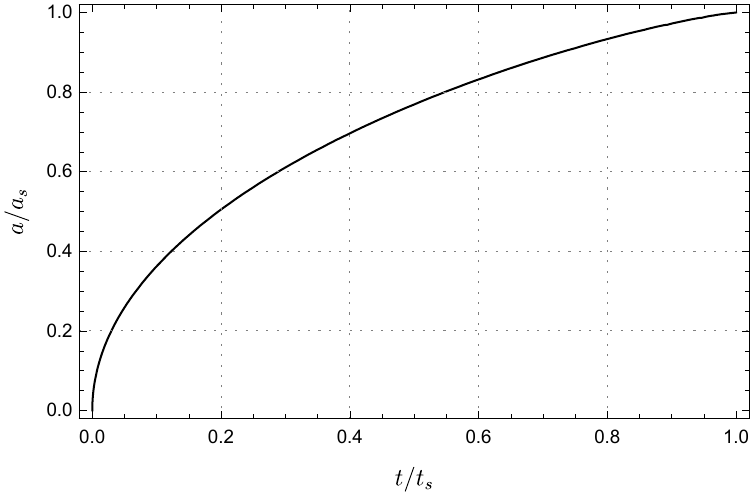}
\caption{The evolution of the decelerating scale factor  (\ref{scalefactor}) in time, normalized by the appropriate values at SFS (here $n=0.2$ and $m=4$). 
\label{fig:evolSF}}
\end{figure}

The dependence of the scale factor $a$ on the cosmological time, $t$, can  be obtained by integrating (\ref{eq:model}), which formally leads to
\begin{align}
\label{scalefactor}
\left(\frac{a}{a_s}\right)^{\frac{m}{2}}{}_2 {\rm F}_1\left(\frac{m}{2},n;1+\frac{m}{2};\frac{a}{a_s}\right)=\frac{\Gamma(1+\frac{m}{2})\Gamma(1-n)}{\Gamma(1+\frac{m}{2}-n)}\frac{t}{t_s},
\end{align}
where ${}_2 {\rm F}_1$ and  $\Gamma$ are the hypergeometric and gamma functions, respectively; see reference \cite{MathFuncBook}. Moreover,
\begin{align}
t_s\coloneqq \frac{a_s^{\frac{m}{2}-n}}{H_\star}\frac{\Gamma(\frac{m}{2})\Gamma(1-n)}{ \Gamma(1+\frac{m}{2}-n)},
\end{align}
is the time of SFS occurrence, which is clearly finite since we focus here on $m\geq0$ and $n<1$.

An inverse of equation (\ref{scalefactor}), i.e. the solution of the scale factor as a function of the cosmological time, $a=a(t)$,  is not possible in terms of elementary functions 
when the parameters $n$ and $m$ are arbitrary. An exception is for $m=2$ (cosmic strings), where one gets
\be
a(t) = a_s \left[ 1 - \left( 1 - \frac{t}{t_s} \right)^{\frac{1}{1-n}} \right] ,
\label{SFSstrings}
\ee
and shows nicely the limits of a BB ($t \to 0$ giving $a \to 0$) and an SFS ($t \to t_s$ giving $a \to a_s$). The relation (\ref{SFSstrings}), applied to (\ref{eq:EoS}), allows to write the behaviour of the varying barotropic index as 
\be
w(t) = - \frac{1}{3} \left(2n+1 \right) + \frac{2}{3} n \left( 1 - \frac{t}{t_s} \right)^{\frac{1}{1-n}} ,
\ee
showing its early negative pressure behaviour and then positive pressure blow-up at $t=t_s$. Out of other solutions, in Fig. \ref{fig:evolSF}, we present the time evolution of the scale factor numerically for $n=0.2$ and $m=4$ as a representative example for the behaviour of our model (\ref{eq:model}).

Two physically relevant regimes at which the inversion of expression (\ref{scalefactor}) is indeed possible in terms of elementary functions are when: (\textit{i}) near the BB initial singularity where $a \approx 0$, 
(\textit{ii}) for $a \approx a_s$ close to the exotic singularity. In the former case, it is possible to obtain that
\be
\frac{a(t)}{a_s} \approx \left(\frac{\Gamma(1+\frac{m}{2})\Gamma(1-n)}{\Gamma(1+\frac{m}{2}-n)}\frac{t}{t_s}\right)^{\frac{2}{m}} \label{asymptBB}
\ee
by expanding (\ref{scalefactor}) in series of small $a/a_s$; see, for instance,  identity 15.1.1 of reference \cite{MathFuncBook}. This indeed reflects the standard fluid behaviour in a universe dominated by radiation ($m=4$) or matter ($m=3$) at small scale factor, as to be expected.
For the later case, i.e. when close to the SFS, the scale factor reads
\begin{align}
\frac{a(t)}{a_s} \approx 1-\left[\frac{\Gamma(\frac{m}{2})\Gamma(2-n)}{\Gamma(1+\frac{m}{2}-n)}\left(1-\frac{t}{t_s}\right)\right]^{\frac{1}{1-n}}, \label{asymptSFS}
\end{align}
see, for instance, identities in section 26.5 of reference \cite{MathFuncBook}.

It is also useful for the incoming discussion to calculate the conformal time $\eta$. This is
\begin{align}
\eta&=\frac{2 a_s^{\frac{m}{2}-1-n}}{(m-2)H_\star} \left(\frac{a}{a_s}\right)^{\frac{m}{2}-1}{}_2 {\rm F}_1\left(\frac{m}{2}-1,n;\frac{m}{2};\frac{a}{a_s}\right)- \eta_0,
\end{align}
where 
\begin{align}\label{eq:eta0}
\eta_0\coloneqq \frac{2 a_s^{\frac{m}{2}-1-n}\Gamma(\frac{m}{2})\Gamma(1-n)}{(m-2)H_\star\Gamma(\frac{m}{2}-n)},
\end{align}
is an integration constant. Please note that for the model (\ref{eq:model}), we have an initial BB at $t=0$ followed by a SFS at $t=t_s$. In terms of the conformal time, this implies that $\eta$ goes from $-\eta_0$ at the BB to 
$0$ at the SFS. Moreover, the conformal time to the BB is finite if and only if $m>2$, which includes the interesting scenarios of radiation ($m=4$) and matter ($m=3$) domination at early times. Using the conformal time, we introduce the particle horizon (PH) and the event horizon (EH) as \cite{Yoshida2023}
\begin{align}
r_{PH}(\eta)=&\,\eta+\eta_0,\label{eq:PH}\\
r_{EH}(\eta)=&-\eta\label{eq:EH}.
\end{align}
Besides, the apparent horizon (AP) in the conformal time reads \cite{Yoshida2023}
\begin{align}\label{eq:AH}
r_{AH}(\eta)\coloneqq \frac{1}{a(\eta)\,H(\eta)},
\end{align}
see, for instance, Ref.~\cite{Melia:2018xtf} and references therein.
We will use these horizons in Section \ref{Penrose} for discussing the causal structure of the corresponding FLRW spacetime.

\subsection{An accelerating towards a finite time pseudo-rip singularity model}

\begin{figure*}
\begin{subfigure}{\columnwidth}
\includegraphics[width=\textwidth]{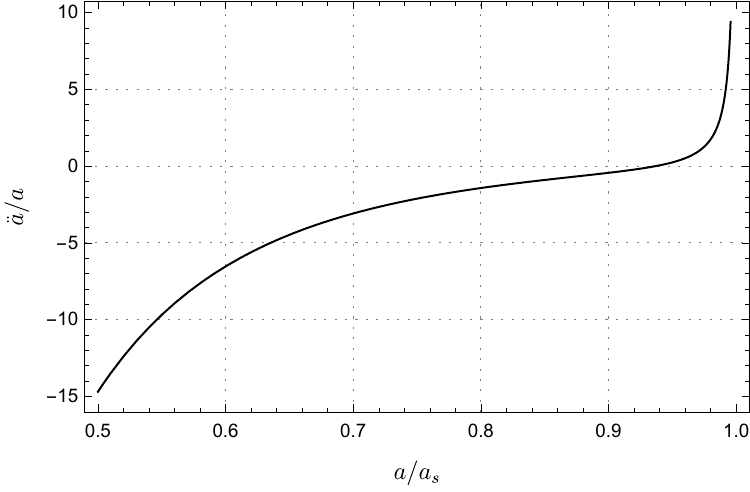}
\end{subfigure}
\hfill
\begin{subfigure}{\columnwidth}
\includegraphics[width=\textwidth]{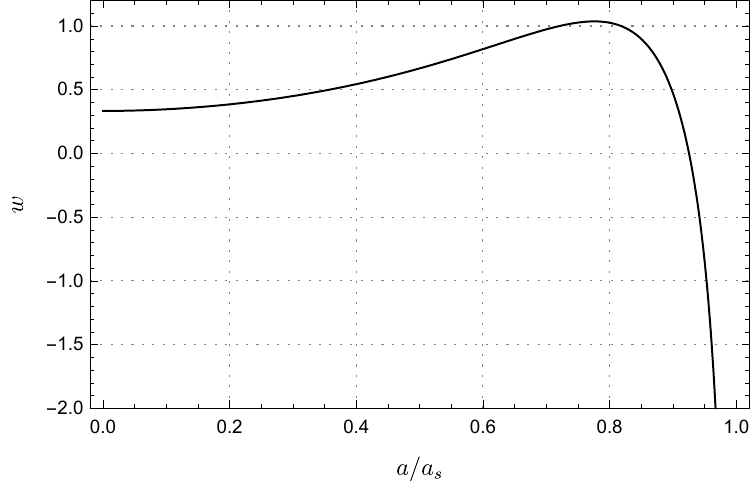}
\end{subfigure}
\caption{
The evolution of the accelerating model (\ref{eq:model2})  for $n=0.2$, $m=4$ (radiation domination at early time) and $H_1a_s^{-2}=H_2a_s^{n}=1$. Left panel: late time evolution of the acceleration $\ddot{a}/a$. Right panel: the evolution of the equation of state parameter $w$. The ''hump'' behaviour of the EoS parameter reaching stiff-fluid is an artefact of using the radiation fluid only (see text and Fig. \ref{fig:weff_LCDM_FTPR})
\label{fig:EoS_mod2}}
\end{figure*}

The new model (\ref{eq:model}) parametrise an ever decelerating universe which lies at total odds with current observations. 
Late-time acceleration which would mimic the dark energy due to a new singularity appearance can be achieved by modifying the model (\ref{eq:model}). An option is to make the terms responsible for the BB singularity ($\sim a^{-\frac{m}{2}}$) and SFS $(\sim (1-a/a_n)^n )$ additive, not multiplicative. This is  
\begin{eqnarray}
\label{eq:model2}
H[a(t)] \equiv \frac{\dot{a}}{a} = \frac{H_1}{a^{ \frac{m}{2}}} - H_2 a_s^{n} \left( 1 - \frac{a}{a_s} \right)^{n},
\end{eqnarray}
where $m$ and $n$ are again positive parameters. Note that minus sign between the two terms is necessary to produce the acceleration. 
Therefore, the second derivative of the scale factor $a(t)$ reads
\bea
\label{2ndmodel2}
\frac{\ddot{a}}{a} 
 &=& \left(1 - \frac{m}{2} \right) H_1^2 a^{-m} + nH_1H_2 a_s^{n-1} a^{1 - \frac{m}{2}} \left(1 -\frac{a}{a_s} \right)^{n-1} \nonumber \\
& + & H_2^2a_s^{2n} \left(1-\frac{a}{a_s}\right)^{2n} - n H_2^2 a a_s^{2n-1} \left(1-\frac{a}{a_s}\right)^{2n-1} \nonumber \\
&-& H_1 H_2 a_s^n a^{- \frac{m}{2}} \left(2 - \frac{m}{2} \right) \left(1-\frac{a}{a_s}\right)^{n} . 
\eea
From (\ref{2ndmodel2}) we notice that the first and fifth terms always decelerates for $m>2$, which includes the case of radiation ($m=4$) and dust ($m=3$) fluids at high redshift.\footnote{Parallel to the discussion below equation (\ref{eq:eta0}), the corresponding $\eta_0$ for the model (\ref{eq:model2}) remains finite when $m>2$.} Conversely, the second term always accelerates and, eventually, blows up at  $a=a_s$ for $0 < n<1$. The third term in (\ref{2ndmodel2}) vanishes at $a=a_s$, whereas the fourth term blows up only if $0 <n<1/2$. Nevertheless, the latter term is always subdominant near the singularity  since the second  term diverges strongly. This behaviour is portrayed in the top of Fig. \ref{fig:EoS_mod2}, where $\ddot a$ changes its sign and, finally, diverges to positive infinity, while $H$ remains finite; see Eq. (\ref{eq:model2}). Contrary to the usual decelerating SFS behaviour discussed before, this signals the presence of an extreme acceleration event at $a_s$ for $n\in(0,1)$. 
	
Moreover, by the very structure of the Hubble rate (\ref{eq:model2}), one can safety claim that the divergence of pressure at $a=a_s$ always takes place in a finite amount of time since the integral
\begin{align}\label{eq:time to FTPR}
t-t_s=\int_0^{a_s} \frac{da}{aH} < \infty 
\end{align}
is clearly finite for positive $m$ and $n$. The evolution of the equation of state parameter $w$ of this model is presented in Fig. \ref{fig:EoS_mod2} for $n=0.2$, \tb{$m=4$} and $H_1a_s^{-2}=H_2a_s^{n}=1$. The universe starts its evolution at $a\to0$ at a BB with the equation of state parameter $w= 1/3$ signalling radiation domination. Then, this parameter has a maximum $w\approx 1.04$ at $a \approx 0.778 a_s$ suggesting an effective stiff-fluid (or even super stiff-fluid) behaviour (kinetic energy domination \cite{khoury2001,turok2002,khoury2002}). After reaching the maximum, the universe turns into an accelerated phase at around $a\approx 0.93 a_s$, and finally it crosses the phantom divide $w=-1$ at $a\approx 0.95 a_s$. This type of behaviour is quite generic for the model (\ref{eq:model2}). Moreover, the lesser the parameter $n$ is in the open interval $(0,1)$, the higher is the maximum of $w$. Indeed, in the extreme case $n = 0.01$, the model exhibits a pronounced peak around $w \approx 12.5$, indicating super-stiff behaviour at $a \approx 0.981 a_s$, shortly before crossing the phantom divide at $a \approx 0.999 a_s$. By contrast, in the opposite limiting case, for instance $n = 0.99$, the maximum is much milder, with $w \approx 0.43$ occurring already halfway to the singularity, i.e. at $a \approx 0.502 a_s$. However, as we will show in Section \ref{sec:LCDM FTPR}, this behaviour is only an artefact of taking into account solely the radiation as the matter content. In a more realistic case of a mimicking $\Lambda$CDM model which introduces the dust fluid, these stiff-fluid maxima disappear.  

In conclusion, the evolution of the model (\ref{eq:model2}) starts at a classical BB singularity effectively dominated by regular fluid with barotropic index $w\sim (m-3)/3$, which generalise the particular scenarios of radiation ($m=4$) and dust matter ($m=3$). Then, it evolves towards a pressure singularity at $a \to a_s$ within a finite amount of cosmic time. The pressure becomes negative and grows unbounded in the vicinity of $a_s$, thus leading to a super-accelerated regime before reaching the curvature singularity. The strong phantom domination at $a \to a_s$ (see Fig.~\ref{fig:EoS_mod2}) closely resembles that found in other phantom-induced singularities such as the BF, the BR, the LR, or the LSBR. Nevertheless, the latter two occur only at infinite future time, whereas the BR leads to $a_s \to \infty$, in contrast with the finite values of $t_s$ and $a_s$ obtained here. The BF does occur at finite values of the cosmic time and scale factor; however, in that case the energy density cannot remain constant. This reasoning also excludes the more general finite-scale-factor, or type III, singularities \cite{Nojiri2005,MPD2014CC}.

According to the usual classification of cosmological singularities solely in terms of the values of the cosmic time, scale factor, (effective) energy density, and (effective) pressure at the singularity \cite{Nojiri2005,MPD2014CC}, one may be tempted to conclude that (\ref{eq:model2}) leads to an SFS or Type II singularity. However, as we argue in the next section, we find this conclusion to be premature, since this model does not satisfy the ECs that served as the very definition and motivation of the original SFS \cite{Barrow2004,Barrow2004a}. Bearing these considerations in mind, we suggest the name ``finite-time pseudo-rip,'' or FTPR in short, for this future singularity. It is important to notice that the FTPR considered here differs from the pseudo-rip originally introduced in Ref.~\cite{PR} in that both $t_s$ and $a_s$ are finite at the FTPR. Moreover, the FTPR represents a true curvature singularity, since the Ricci scalar clearly diverges when the effective pressure blows up, in contrast with the pseudo-rip scenario. Alike the SFS, geodesics could, in principle, be extended through the FTPR as $a$ and $\dot{a}$ remain regular, whereas only $\ddot{a}$ diverges, leading to potentially severe tidal forces acting on extended objects \cite{Leonardo}. We comment more on this in section \ref{ECond}.

\section{Energy conditions for new models}
\label{ECond} 


In this section we discuss the ECs of Hawking and Penrose \cite{HE}  for the isotropic (flat) FLRW cosmological models presented before. These conditions can be summarised as follows:
\begin{itemize}[leftmargin=+5mm]
\item Null Energy Condition (NEC): $\varrho c^2 + p \geq 0$;
\item Weak Energy Condition (WEC): $\varrho c^2 \geq  0$,  $\varrho c^2 + p \geq 0$;
\item Dominant energy Condition (DEC): $\varrho c^2 \geq \mid p \mid$;
\item Strong Energy Condition (SEC): $\varrho c^2 + p \geq 0$, $\varrho c^2 + 3p \geq 0$.
\end{itemize}
Generally the DEC implies the WEC and, then, the NEC. While the SEC implies the NEC, which means that NEC is the most extreme violation of all. 

Applying the equations (\ref{rho1}) and (\ref{p1}) one can write the above conditions in the form: 
\begin{itemize}[leftmargin=+5mm]
\item NEC: $\varrho c^2 + p = - \frac{c^2}{4 \pi G} \dot{H} \geq 0$;
\item WEC: $\varrho c^2 = \frac{3 c^2}{8 \pi G}H^2  \geq 0$,  $\varrho c^2 + p = - \frac{c^2}{4 \pi G} \dot{H} \geq 0$;
\item DEC-I: $\varrho c^2 \geq -p  \Rrightarrow - \frac{c^2}{4 \pi G} \dot{H} \geq 0$;
\item DEC-II: $\varrho c^2 \geq  p\Rrightarrow \dot{H} + 3H^2 \geq 0$; 
\item SEC: $\varrho c^2 + 3p \geq 0 \Rrightarrow \frac{\ddot{a}}{a} = \dot{H} + H^2 \leq 0$;
\end{itemize} 
where the DEC has been divided into DEC-I and DEC-II depending whether the pressure takes negative or positive values, respectively. In this approach, the DEC is satisfied only if both DEC-I and DEC-II hold.

\subsection{Decelerating SFS model}

For the decelerating model (\ref{eq:model}), using (\ref{eq:EoS}), the NEC reads 
\be
\varrho c^2 + p = \varrho c^2 \left(\frac{m}{3} + \frac{2n}{3} \frac{a}{a_s - a} \right) \geq 0 , \label{NEC1}
\ee
and is always fulfilled since we have assumed that $m \geq 0$ and $n \geq 0$. Of course at singularity for $a=a_s$, $\varrho c^2 + p \to \infty$, because of the pressure blow-up. The WEC is also trivially fulfilled on the base of (\ref{rho1}) ($\varrho \sim H^2 \geq 0$ always), and the condition (\ref{NEC1}). 

In order to check the expected violation of DEC at the SFS, we inspect separately the DEC-I and DEC-II. Imposing the DEC-I implies the NEC, which is satisfied for all $a\in[0,a_s]$ and for the parameter ranges of interest, i.e. $n\in(0,1/2)$ and $m\geq0$. On the contrary, satisfying DEC-II requires
\begin{eqnarray}
   \left( 3-\frac{m}{2}\right)\left(1-\frac{a}{a_s}\right) -n \frac{a}{a_s}\geq 0,
\end{eqnarray}
which can only be fulfilled at early stages in the evolution ($a/a_s \approx 0$) of the model provided that $0\leq m\leq6$. This includes, for instance, the case in which radiation ($m=4$) or dust ($m=3$) dominates at early times. However, this condition is always violated when $a \approx a_s$ and, consequently, so is the DEC. This is to be expected, since the SFS should violate the DEC \cite{Barrow2004,Barrow2004a}.

Finally, the SEC fulfilment demands 
\begin{eqnarray}
   \left( 1-\frac{m}{2}\right)\left(1-\frac{a}{a_s}\right) - n \frac{a}{a_s}\leq 0,
\end{eqnarray}
which requires $m\geq2$ to hold. Since positive $n$ always contributes to deceleration in this model, this additional condition for the SEC (i.e. $m\geq2$) can be interpreted as enforcing $w\geq -1/3$ in the limit $a\to0$.

Summarising the ECs for SFS model (\ref{eq:model}), the NEC and WEC are always fulfilled. The SEC demands the extra condition $m\geq2$ to hold. However, the DEC is always violated (at least) near the SFS. The singularity is weak, at least in the sense of Tipler \cite{Tipler} and Kr\'olak \cite{Krolak}, though the pressure has a finite-time divergence. So, SFS can also be called ``sudden pressure singularity" since the energy density does not blow-up. 

\subsection{Accelerating FTPR model}

For the model (\ref{eq:model2}), the NEC (i.e. $\dot H \leq0$) reads
\bea
\label{NEC2}
&&\frac{m}{2} H_1 H_2 a_s^{n} a^{-\frac{m}{2}}  \left(1-\frac{a}{a_s}\right)^{n} - n H_2^2 a a_s^{2n-1} \left(1-\frac{a}{a_s}\right)^{2n-1} \nonumber  \\
&&-\frac{m}{2} \frac{H_1^2}{a^4} +nH_1H_2 a_s^{n-1} a^{1-\frac{m}{2}} \left(1 -  \frac{a}{a_s} \right)^{n-1} \leq 0, 
\eea
which shows that it is fulfilled at $t=0$ when $a\approx0$, i.e. at the BB, since the third term in (\ref{NEC2}) dominates and it is negative. However, at $t\approx t_s$ when $a \approx a_s$ we can differentiate two scenarios. These are: \textit{i}) $0 \leq n < 1/2$, and \textit{ii}) $1/2 < n < 1$. In the former case, the second term in (\ref{NEC2}) dominates over the fourth term, since the power in the blowing up term is more negative (i.e. $n-1 < 2n-1 <0$), though both of them blow up at $a=a_s$. The second term contributes positively while the fourth negatively and so we conclude that NEC is violated, which in view of the finite value of the energy density given by (\ref{eq:model2}) seems to fulfil the condition of a pseudo-rip-like singularity \cite{PR1}. In scenario \textit{ii}), only the second term in (\ref{NEC2}) diverges, while the fourth term vanishes at $a=a_s$ and so again we have a pseudo-rip-like singularity. Bearing in mind that WEC requires additionally only the positivity of the energy density, which is always fulfilled according to (\ref{rho1}), we can conclude that the WEC is also violated at $a\approx a_s$ due to the second term in (\ref{NEC2}). 

As for DEC-II we have from (\ref{eq:model2}) and (\ref{NEC2}) that 
\bea
\label{DEC-II}
&&   \left(3 - \frac{m}{2} \right) \frac{H_1^2}{a^{m}} +  \left(\frac{m}{2} - 6 \right) H_1 H_2 a_s^{n} a^{-\frac{m}{2}}  \left(1-\frac{a}{a_s}\right)^{n}  \nonumber \\
&& + nH_1H_2 a_s^{n-1} a^{1-\frac{m}{2}} \left(1 -  \frac{a}{a_s} \right)^{n-1}  + 3H_2^2 a_s^{2n}  \left(1-\frac{a}{a_s}\right)^{2n} \nonumber \\
&&  -  n H_2^2 a a_s^{2n-1} \left(1-\frac{a}{a_s}\right)^{2n-1}  \geq 0. 
\eea
Parallel to the discussion for the previous model, this condition is fulfilled at BB singularity where $a(0) = 0$ if $0\leq m\leq6$. As for $a = a_s$, unlike in the previous case, now the third term with the power $(n-1)$ contributes positively securing the fulfilment of DEC-II. The DEC-I, however, cannot be satisfied at $a\approx a_s$ since this condition is directly related to the NEC. Thus, the DEC-I and, therefore, the DEC are indeed violated for (\ref{eq:model2}) at least at $a\approx a_s$. 

The SEC can be discussed while looking at the formula (\ref{2ndmodel2}) with the conclusion that it is fulfilled at BB singularity for $m\geq2$ (the first term $\propto a^{-m}$ contributes negatively),  while it is not at $a=a_s$ since the dominating second term with the power $(n-1)$ contributes positively.

In summary, for the model (\ref{eq:model2}), all the aforementioned ECs are satisfied at the onset of the corresponding evolution provided that $2\leq m\leq6$. However, each of them is violated at some stage during the expansion history. Consequently, at $a \approx a_s$ none of the ECs holds. These violations are, of course, associated with the term proportional to the power $n-1$ in expression (\ref{NEC2}), which induces the phantom crossing and ultimately drives the effective pressure to diverge to minus infinity. This behaviour stands in sharp contrast to that of the SFS model (\ref{eq:model}), where the effective pressure diverges to positive infinity and, as a result, only the DEC is violated at $a\to a_s$. This indicates that the event occurring at $a \to a_s$ in model (\ref{eq:model2}) is fundamentally different from that of the specific SFS model (\ref{eq:model}) considered here, as well as from the general definition of SFS \cite{Barrow2004}. For this reason, we proposed the designation FTPR for the future singularity arising in model (\ref{eq:model2}), in order to clearly distinguish it from the SFS case, in which only the DEC is violated.

\subsection{Geodesics and tidal forces}
\label{geodesics}

The studies of geodesic deviation equation (see for example \cite{Keresztes2009,Keresztes2010})  show that the acceleration of geodesic deviation vector reads
\be 
\dot{u}^a = - R^{a}_{cbd} \eta^b u^c u^d ,
\label{acc}
\ee
where $ R^{a}_{cbd}$ are the components of the Riemann tensor for the FLRW metric ($a, b = 0, 1, 2, 3$ or $t, r, \theta, \phi$) which lead to field equations (\ref{rho1}) and (\ref{p1}), $\eta^b$ is the geodesic deviation vector, and $u^c$ is the 4-velocity vector. Since the 4-velocity for FLRW model has just one non-zero component, which is $u^t = -1$, then the acceleration of the deviation (\ref{acc}) reads
\be
\dot{u}^a = - R^{a}_{tbt} \eta^b \propto \ddot{a}  ,
\label{acc}
\ee
because the only non-vanishing components of the Riemann tensor are $R_{trtr} = - \ddot{a} a$ and $R_{t \phi t \phi} = R_{t \theta t \theta} = - \ddot{a} a r^2 \sin{\theta}$. This means that the tidal forces at both SFS and FTPR reach infinite values as both expressions for $\ddot{a}$ in (\ref{eq:ddot a}) and (\ref{2ndmodel2}) diverge for $t\to t_s$, eventually stopping the deviation of geodesics and allowing the extension of them beyond singularity. On the other hand, the velocity of deviation is 
\be
v^a = u^b \nabla_b \eta^a \propto H = \frac{\dot{a}}{a} ,
\ee
reaching a constant values at $t=t_s$ as it can be seen from our defining models expressions (\ref{eq:model}) and (\ref{eq:model2}). 

	
\section{Singularities and horizons}
\label{Penrose}
	
In order to study the behaviour of the cosmological horizons (\ref{eq:PH})-(\ref{eq:AH}) we build the Penrose diagrams \cite{HE} related to new models (\ref{eq:model}) and (\ref{eq:model2}). For this, we start with the isotropic and spatially flat ($k=0$) FLRW line element written down in conformal time as 
\begin{align}\label{eq:lineE}
ds^2=a(\eta)^2\left(-d\eta^2+dr^2+r^2d\Omega_3^2\right),
\end{align}	
where $\eta\in[-\eta_0,0]$, $\eta_0>0$, $r\geq0$, and $d\Omega_3$ being the line element of a 2-dimensional sphere.
Introducing the null coordinates
\begin{align}
u=&\eta-r,\\
v=&\eta+r, 
\end{align}
where $u\in(-\infty,0]$ and $v\in[-\eta_0,\infty)$ with $u\leq v$, we may bring the line element (\ref{eq:lineE}) into 
\begin{align}
ds^2=a(u,v)^2\left[-dudv+\frac{(v-u)^2}{4}d\Omega_3^2\right].
\end{align}

\begin{table*}
\setlength\arrayrulewidth{0.6pt}
\setlength{\tabcolsep}{3.mm}
\renewcommand{\arraystretch}{1.5}
\begin{tabular}{c  c  c  c  c  c  c  c  c  c }
\toprule 
Name & $t$ & $\eta$ & $r$ & $u$ & $v$ & $U$ & $V$ & $T$ & $R$ \\
\midrule
$T_0$ & $0$ & $-\eta_0$ & $0$ & $-\eta_0$ & $-\eta_0$ & $-\arctan(\eta_0)$ & $-\arctan (\eta_0)$ & $-2\arctan(\eta_0)$ & $0$ \\
$i^0$ & $0$ & $-\eta_0$ & $\infty$ & $-\infty$ & $\infty$ & $-\frac{\pi}{2}$ & $\frac{\pi}{2}$ & $0$ & $\pi$ \\
$O$  & $t_s$  & $0$  & $0$  & $0$  & $0$  & $0$  & $0$ & $0$ & $0$ \\
$i^0$ & $t_s$ & $0$ & $\infty$ & $-\infty$ & $\infty$ & $-\frac{\pi}{2}$ & $\frac{\pi}{2}$ & $0$ & $\pi$ \\
\bottomrule
\label{tab:points}
\end{tabular}
\caption{Relevant points for the Penrose diagram in the different sets of coordinates used in Sec. \ref{Penrose}.\label{tab:points} }
\end{table*}

\noindent The coordinates $(u,v)$ can be transformed into the new ones, $(U,V)$ by the following compactification scheme
\begin{align}
U=& \arctan(u),\\
V=&\arctan(v),
\end{align}
which fulfil that $U\in(-\pi/2,0]$ and $V\in[-\arctan(\eta_0),\pi/2)$, and $U\leq V$. In these coordinates, the line element reads
\begin{align}
ds^2=\frac{a(u,v)^2}{\cos^2U\cos^2V}\left[-dUdV+\frac{\sin^2(v-u)}{4}d\Omega_3^2\right].
\end{align}
	
Finally, this line element can be mapped into the Einstein static cylinder universe by considering the following time- and space-like compactified coordinates
\begin{align}
T=&U+V,\\
R=&V-U.
\end{align}
Note that $T\in[-T_0,0]$ and $R\in[0,\pi)$ where
\begin{align}
T_0\coloneqq -2\arctan(\eta_0).
\end{align}
Moreover, $T_0>-\pi$ since $\eta_0$ is finite for the models of our interest (recall that generally $m>2$ and $0<n<1$). The line element now reads
\begin{align}
ds^2=\frac{a(T,R)^2}{\left[\cos(T)+\cos(R)\right]^2}\left(-dT^2+dR^2+\sin^2(R) d\Omega_3^2\right),
\end{align}
which is nothing, but the Einstein cylinder up-to a conformal factor.

\begin{figure*}
\begin{subfigure}{\columnwidth}
\includegraphics[width=\textwidth]{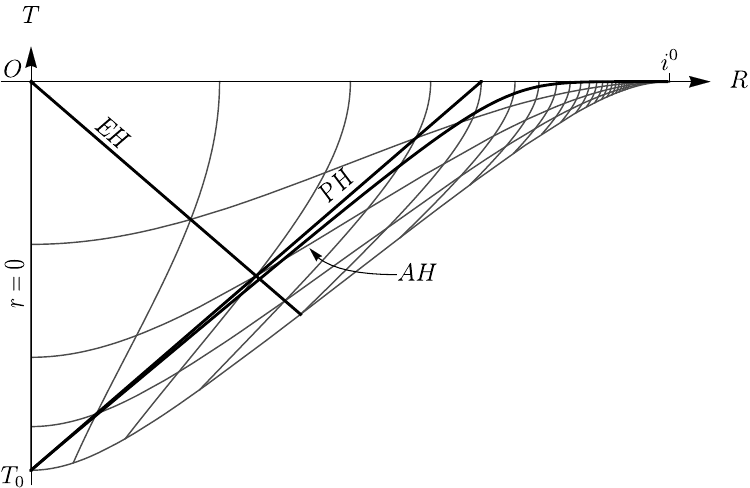}
\end{subfigure}
\hfill
\begin{subfigure}{\columnwidth}
\includegraphics[width=\textwidth]{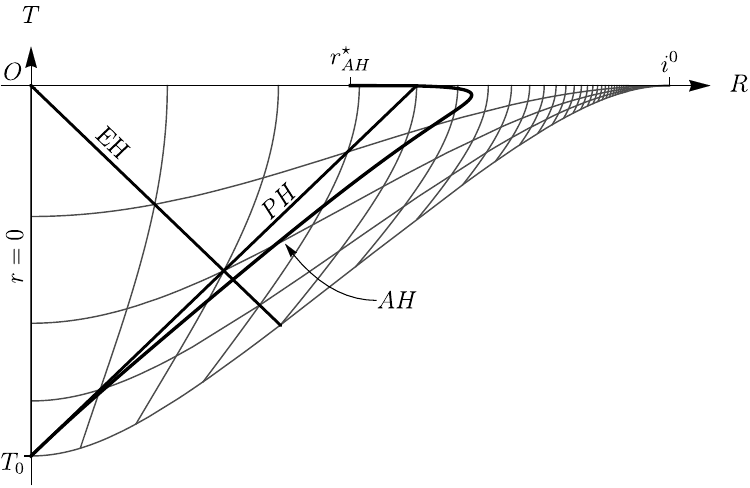}
\end{subfigure}
\caption{
Comparison of the Penrose diagrams for the models discussed in this section, where EH stands for the event horizon, PH for the particle horizon, and AH is the apparent horizon. Left panel: diagram for the model (\ref{eq:model}) representing a decelerating SFS with $n=0.2$ and $m=4$ (radiation domination at $a\approx0$). Right panel: diagram for the accelerating model (\ref{eq:model2}) with $n=0.2$, $m=4$ (radiation domination at $a\approx0$), $a_s=1$ and $H_1=H_2=1$. In addition, $r_{AH}^\star$ represents the spatial radius of AH evaluated at the FTPR.\label{fig:penrose}}
\end{figure*}

The important points for the Penrose diagrams are summarised in Table \ref{tab:points}.  It is interesting to note that the past time-like infinity point $i^-$ and  null past infinity $\mathcal{I}^-$ are not 
present in Table \ref{tab:points} since $\eta$ in our models is always finite when $m>2$ and $0<n<1$. Examples of the Penrose diagrams for the decelerating SFS (\ref{eq:model}) and accelerating FTPR model (\ref{eq:model2}) can be found in Fig. \ref{fig:penrose}. In principle, both diagrams can be extended beyond the upper boundary $T = 0$, which corresponds to the occurrence of these singularities, since these events are geodesically extendible \cite{Leonardo}. Similar diagrams for different SFS models have been previously found in Ref. \cite{GRG2018}. 	

For the decelerating SFS modelled by equation (\ref{eq:model}), the spatial radius of the different horizons previously mentioned obey the following constraints 
\begin{align}
&	r_{EH}>r_{AH}\geq r_{PH} \ \ \textup{at BB,}\\
&   r_{AH}>r_{PH}>r_{EH} \ \ \textup{at SFS,}
\end{align}
see left panel in Fig. \ref{fig:penrose}. However, this spatial hierarchy between different horizons changes when having an accelerated regime, such as in model (\ref{eq:model2}) (see right panel in Fig. \ref{fig:penrose}). More concretely, the spatial radius for the AH grows more slowly in the accelerating FTPR universe than in the decelerating SFS one. This should not be surprising given that
\begin{align}
\diff{\,r_{AH}}{\eta}=-\frac{\ddot a}{aH^2},
\end{align}
whereas,
\begin{align}
\diff{\,r_{PH}}{\eta}&=1,\\
\diff{\,r_{EH}}{\eta}&=-1,
\end{align}		
see definitions in equations (\ref{eq:PH})-(\ref{eq:AH}). Consequently, the presence of an accelerating/decelerating regime directly influences the (conformal) growth rate of the AH, but not that of the PH and EH. In adition, $r_{AH}(t_s)=r_{AH}^\star<\infty$ in the FTPR model (\ref{eq:model2}), whereas it diverges for the SFS (\ref{eq:model}); see Fig. \ref{fig:penrose}.


\section{Raychaudhuri averaging and related strength of exotic singularities} 
\label{Raych}

The general SFS scenario and the FTPR found here are weak singularities \cite{Nojiri2005} according to the definitions of Tipler \cite{Tipler} and Kr\'olak \cite{Krolak}, and therefore extendible \cite{Leonardo}. A different definition of singularity criterion was given by Raychaudhuri \cite{raych98}, and later developed by one of the Authors \cite{PLB2011}. According to these considerations, one is always able to take an average of any physical or kinematic scalar quantity $\chi$ over the entire (open)
spacetime in the form
\be
<\chi> \equiv \left[ \frac{\int_{-x_0}^{x_0} \ldots \int_{-x_3}^{x_3} \chi \sqrt{-g} \,{\rm d}^4 x}{\int_{-x_0}^{x_0} \ldots \int_{-x_3}^{x_3}
\sqrt{-g} \,{\rm d}^4 x}\right]_{\lim_{x_0, \ldots x_3 \to \infty}}.
\label{averchi}
\ee
For a FLRW cosmological model, we focus on averaging the acceleration scalar $\chi = \theta_{,\mu}u^{\mu} = 3H^2 (q+1)$, where $\theta$ is the expansion scalar, $q = - \ddot{a}a/\dot{a}^2$ is the deceleration parameter, $\mu, \nu = 0, 1, 2, 3$. In our case, instead of infinite limit of the coordinates, we will limit ourselves to averaging on the domain of applicability of the model. Then, according to (\ref{averchi})  \cite{PLB2011}
\be
<\dot{\theta}> = \lim_{\genfrac{}{}{0pt}{}{t_0 \to 0}{t_1 \to t_s}} \frac{3\int_{t_0}^{t_1}a^3\left(\frac{\ddot{a}}{a} - \frac{\dot{a}^2}{a^2} \right){\rm d}t}{\int_{t_0}^{t_1}a^3{\rm d}t}.
\label{thetaav}
\ee

For our decelerating model (\ref{eq:model}), on the one hand, the denominator of (\ref{thetaav}) gives 
\begin{align}
\lim_{\genfrac{}{}{0pt}{}{t_0 \to 0}{t_1 \to t_s}}&\int_{t_0}^{t_1} a^3{\rm d}t = \frac{a_s^{3-n+\frac{m}{2}}\Gamma(3+\frac{m}{2})\Gamma(1-n) }{H_\star \Gamma(4+\frac{m}{2}-n)}, 
\end{align}
which is finite for the values of $m$ and $n$ considered here. The numerator of (\ref{thetaav}), on the other hand, resolves in 
\begin{align}
\lim_{\genfrac{}{}{0pt}{}{t_0 \to 0}{t_1 \to t_s}}3\int_{t_0}^{t_s}a^3\left(\frac{\ddot{a}}{a} - \frac{\dot{a}^2}{a^2} \right){\rm d}t=  -\frac{9a_s^{3+n-\frac{m}{2}}H_\star \Gamma(1+n)}{\Gamma(4+n-\frac{m}{2})}, 
\end{align}
by using both (\ref{eq:model}) and (\ref{eq:ddot a}). Clearly this contribution is also finite for the values of $m$ and $n$ considered here. Therefore, the SFS which appears in the decelerating model (\ref{eq:model}) is weak in the sense of Raychaudhuri spacetime averaging (cf. \cite{PLB2011}). 

For the accelerating model (\ref{eq:model2}), the denominator in the Raychaudhuri averaging (\ref{thetaav}) remains finite and positive since the co-moving 3-volume is a smooth positive-defined function on that compact integration range. For the numerator, expressions (\ref{eq:model2}) and (\ref{2ndmodel2}) lead to 
\begin{align}\label{eq:RayFTPR}
    \lim_{\genfrac{}{}{0pt}{}{t_0 \to 0}{t_1 \to t_s}}3&\int_{t_0}^{t_s}a^3\left(\frac{\ddot{a}}{a} - \frac{\dot{a}^2}{a^2} \right){\rm d}t \nonumber\\
    =&-\frac{3m}{6-m}a_s^{3-\frac{m}{2}}H_1\left(1-\lim_{a_0\to0}\left(\frac{a_0}{a_s}\right)^{3-\frac{m}{2}}\right)\nonumber\\
    &+\frac{18a_s^{n+3}H_2}{(3+n)(2+n)(1+n)},
\end{align}
which is finite provided that $0\leq m<6$.\footnote{The case of $m=6$ cannot be directly evaluated using expression (\ref{eq:RayFTPR}). When properly addressed, this case introduces a logarithm divergence at $a_0\to0$.} Note that this condition is clearly satisfied in the physically relevant scenarios, in particular, for radiation ($m=4$) and dust ($m=3$). This shows that the FTPR singularity appearing in the accelerating model (\ref{eq:model2}) is also weak in the sense of Raychaudhuri spacetime averaging.

\section{Embedding the FTPR in $\Lambda$CDM cosmology\label{sec:LCDM FTPR}}

So far, we have introduced the FTPR as a new type of cosmological singularity and discussed its differences with the usual SFS and PR scenarios. In this section, we elaborate on how the FTPR can arise within a cosmology framework more closely related to $\Lambda$CDM.

Taking into account the structure of the FTPR model (\ref{eq:model2}), we propose the following $\Lambda$CDM-like expansion profile 
 \begin{equation}\label{eq:model LCDM FTPR}
    H^2(a)= \mu_1 a^{-4} + \mu_2a^{-3}+\mu_3\left[1-n\left(1-\frac{a}{a_s}\right)^{2n}\right],
\end{equation}
where $a_s$ is the value of the scale factor at the FTPR and $\mu_i$ are positive constants. These constants can be re-expressed in the more familiar form
\begin{eqnarray}
    H_0^2 \Omega^{(0)}_{\rm rad}&=&\mu_1,\\
    H_0^2 \Omega^{(0)}_{\rm mat}&=&\mu_2,\\
    H_0^2 \Omega^{(0)}_{\rm DE}&=&\mu_3\left[1-n\left(1-\frac{1}{a_s}\right)^n\right],\label{eq:mu3}
\end{eqnarray}
being $H_0$ the current value of the Hubble rate at $a=1$.

From the expansion rate (\ref{eq:model LCDM FTPR}) it follows that
\begin{align}\label{eq:dotH}
    \dot H (a) =& -2 \mu_1 a^{-4}-\frac32\mu_2 a^{-3}\nonumber\\
    &+n^2 \mu_3\left[\left(1-\frac{a}{a_s}\right)^{2n-1}-\left(1-\frac{a}{a_s}\right)^{2n}\right],
\end{align}
and, therefore, the cosmic acceleration reads
\begin{align}
    &\frac{\ddot a}{a}= -\mu_1 a^{-4}-\frac12 \mu_2 a^{-3}\\
    &+\mu_3\left[1+n^2\left(1-\frac{a}{a_s}\right)^{2n-1}-n(n+1)\left(1-\frac{a}{a_s}\right)^{2n}\right].\nonumber
\end{align}
Hence, for $0<n<1/2$ the Hubble rate converges to a constant value as $a\to a_s$, while its cosmic time derivative diverges to infinity. This behaviour corresponds to a pressure singularity $p\to -\infty$ at $a\to a_s$. Moreover, for a monotonically expanding universe this divergence is clearly reached within a finite amount of cosmic time; see equation (\ref{eq:time to FTPR}). Thus, the evolution leads to the appearance of a FTPR for $0<n<1/2$.

Since the cosmic time derivative of the Hubble rate is negative during (at least) radiation and matter domination but positive near the FTPR, let $a_\star$ denote the (future) moment at which $\dot H$ vanishes. Then, for $a>a_\star$ the NEC and, consequently, all ECs are violated, as discussed in Section \ref{ECond}.  The Raychaudhuri averaging can also be evaluated for this model, although the integral cannot be expressed in terms of simple mathematical functions as in the previous examples. To that end, note that the denominator in expression (\ref{thetaav}) is always finite for a monotonically increasing function $a(t)$ integrated over a compact domain. The numerator in (\ref{thetaav}) can conveniently be re-expressed as
\begin{equation}\label{eq:rayN}
    3\int_{t_0}^{t_1}a^3\dot H {\rm d}t = 3\int_{a_0}^{a_\star}\frac{a^2\dot H }{H}{\rm d}a+3\int_{a_\star}^{a_1}\frac{a^2\dot H}{H}{\rm d}a,
\end{equation}
where $\dot H(a_\star)=0$. The first integral has a proper limit as $a_0\to 0$, since the quotient $a^2\dot H/H$ is well defined at vanishing scale factor. Thus, this integral contributes with a finite quantity to the averaging. Conversely, the second integral has a potentially ill-defined limit as $a_1\to a_s$, since $\dot H$ diverges at the FTPR. However, it can be shown that there exists a function $f(a)$ with a finite integral such that $a^2\dot H/H\leq f(a)$ for $a\in[a_\star, a_s]$. Indeed,
\begin{equation}
    f(a)= \mu\left(1-\frac{a}{a_s}\right)^{2n-1+\epsilon},
\end{equation}
with $\mu$ some constant and $\epsilon\ll1$ fulfils both conditions. Hence, the second integral in expression (\ref{eq:rayN}) has a finite limit at $a\to a_s$ and, therefore, the Raychaudhuri averaging of the FTPR model (\ref{eq:model LCDM FTPR}) remains finite.

\begin{figure}
\hspace*{-0.5cm}
\begin{subfigure}{0.49\textwidth}
\includegraphics[width=\textwidth]{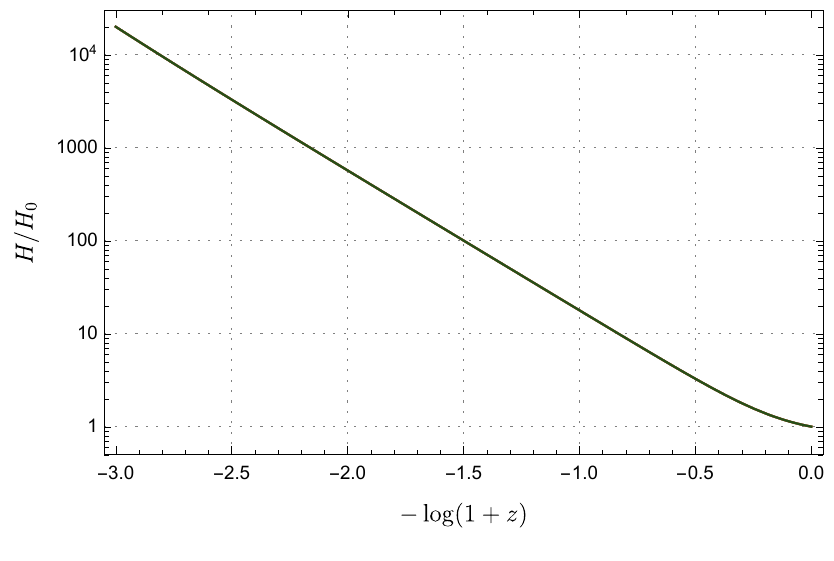}
\end{subfigure}
\begin{subfigure}{0.49\textwidth}
\hspace*{-0.5cm}
\includegraphics[width=\textwidth]{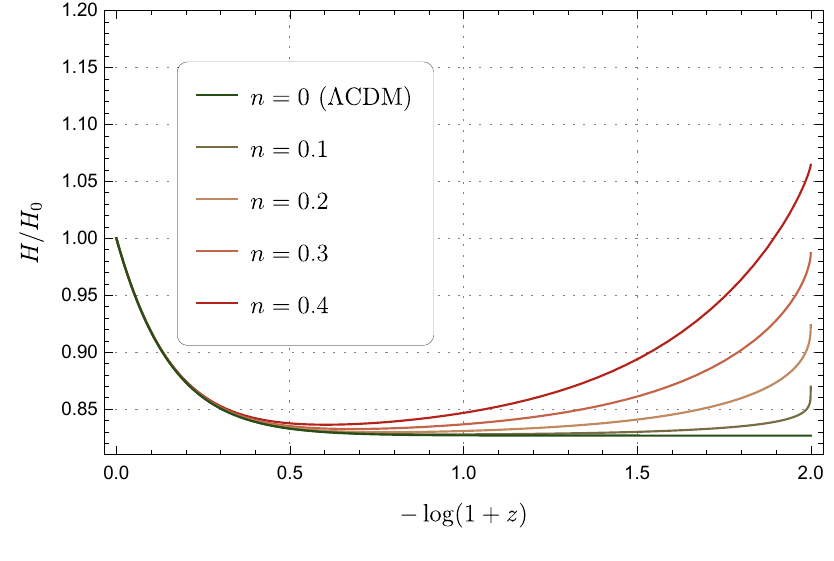}
\end{subfigure}
\caption{The evolution of the Hubble rate in model (\ref{eq:model LCDM FTPR}) for different values of $0<n<1/2$ and $a_s=100$. The initial conditions are selected according to the Planck data \cite{Planck:2018vyg} $\Omega_{\rm rad}^{(0)}\approx8.4\cdot 10^{-5}$, $\Omega_{\rm mat}^{(0)}\approx0.32$ and $\Omega_{\rm DE}^{(0)}\approx0.68$. Top panel: past expansion history. Bottom panel: future expansion profile.}
\label{fig:H_LCDM_FTPR}
\end{figure}

\begin{figure*}
\centering
\hspace*{-0.5cm}
\begin{subfigure}{0.49\textwidth}
\includegraphics[width=\textwidth]{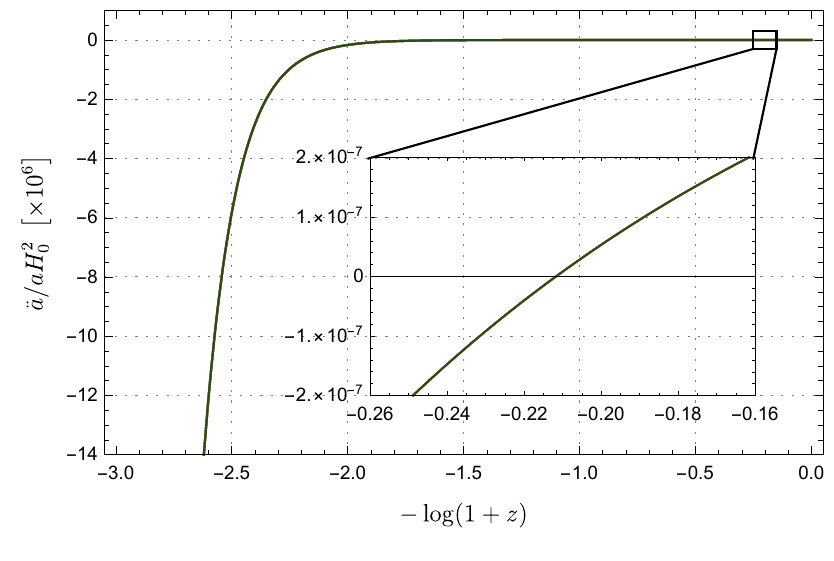}
\end{subfigure}
\hspace*{0cm}
\begin{subfigure}{0.49\textwidth}
\includegraphics[width=\textwidth]{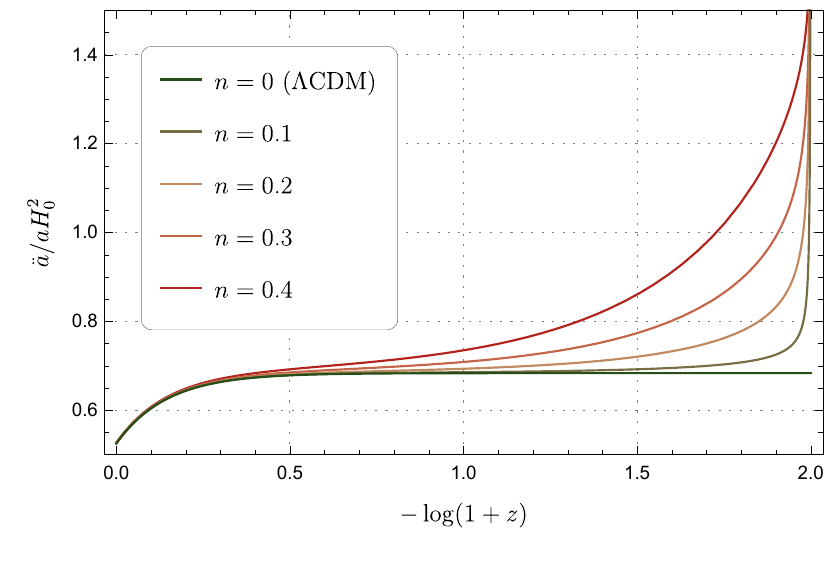}
\end{subfigure}
\caption{Comparison of the acceleration profile of our model (\ref{eq:model LCDM FTPR}) for different values of $n$ within the range $0< n< 1/2$ with that of $\Lambda$CDM. The figures are generated by setting $a_s=100$ and using the Planck \cite{Planck:2018vyg} data for the initial conditions: $\Omega_{\rm rad}^{(0)}\approx8.4\cdot 10^{-5}$, $\Omega_{\rm mat}^{(0)}\approx0.32$ and $\Omega_{\rm DE}^{(0)}\approx0.68$. Left panel: the past acceleration profile where different values of $n$ produce indistinguishable results from $\Lambda$CDM. Right panel: the future acceleration profile.}
\label{fig:acc_LCDM_FTPR}
\end{figure*}

\begin{figure}
\hspace*{-0.5cm}
\includegraphics[width=0.49\textwidth]{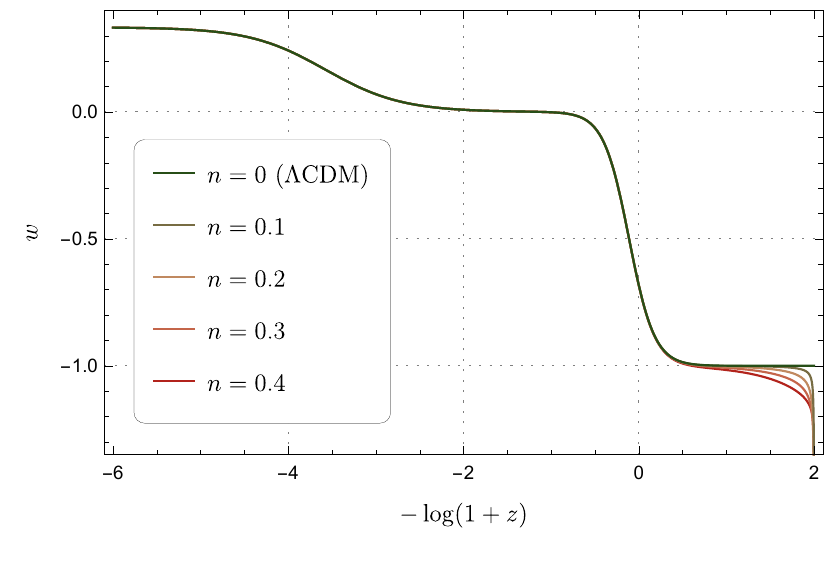}
\caption{Evolution of the effective equation of state parameter for the total fluid on the r.h.s. of expression (\ref{eq:model LCDM FTPR}) for different values of $n$ and $a_s=100$. Initial conditions are selected according to the Planck data \cite{Planck:2018vyg} $\Omega_{\rm rad}^{(0)}\approx8.4\cdot 10^{-5}$, $\Omega_{\rm mat}^{(0)}\approx0.32$ and $\Omega_{\rm DE}^{(0)}\approx0.68$. }
\label{fig:weff_LCDM_FTPR}
\end{figure}

It is also worth noting that, owing to the structure of the model (\ref{eq:model LCDM FTPR}), the past background expansion becomes practically indistinguishable from that of $\Lambda$CDM for fixed present-day energy densities $\Omega_i^{(0)}$ when $a_s\gg1$ and/or $0<n\ll1$. This behaviour can be seen in Figs.~\ref{fig:H_LCDM_FTPR} to \ref{fig:weff_LCDM_FTPR}, where we show the background evolution of the model for $a_s=100$. It is important to emphasise that these figures have been generated by taking the Planck data \cite{Planck:2018vyg} $\Omega_{\rm rad}^{(0)}\approx8.4\cdot 10^{-5}$, $\Omega_{\rm mat}^{(0)}\approx0.32$ and $\Omega_{\rm DE}^{(0)}\approx0.68$ at face value. Therefore, they do not represent a proper confrontation of our model with observational data, but rather an exploration of its quantitative behaviour for a set of reasonable initial conditions.
Fig. \ref{fig:H_LCDM_FTPR} shows that the past background expansion closely follows that of $\Lambda$CDM, whereas in the future $H(a)$ becomes an increasing function of time. Likewise, the past cosmological acceleration profile for $0<n<1/2$ strongly resembles that of the standard cosmological model; see Fig. \ref{fig:acc_LCDM_FTPR}. However, in the future the acceleration exceeds that produced by a cosmological constant and eventually diverges as $a \to a_s$. Furthermore, Fig. \ref{fig:weff_LCDM_FTPR} confirms the standard radiation- and matter-dominated epochs, showing that the value of $n$ has no significant effect on their duration or on the transition points between eras. This figure also illustrates the violation of the NEC in the future and the sudden divergence of the pressure to negative infinity at the FTPR for $0<n<1/2$.

It is also important to emphasize that for $a_s \gg 1$ and/or $0 < n \ll 1$, the model closely mimics $\Lambda$CDM at the background level, as the FTPR term effectively behaves as a cosmological constant in the past and present. In consequence, the parameters $n$ and $a_s$ cannot individually be constrained by current observational data. However, a full treatment of cosmological perturbations remains to be performed for this model, which may reveal significant deviations from the standard picture. Future observations sensitive to the phantom regime epoch (where $a \sim 1$) or to the evolution of the equation of state parameter also could, in principle, help break this degeneracy

As an extra comment, we mention that the stiff-fluid behaviour of the EoS parameter $w$ seen in the less realistic radiation only model (cf. Fig. \ref{fig:EoS}) has disappeared due to the admittance of standard cosmological dust fluid (cf. Fig. \ref{fig:weff_LCDM_FTPR}).

\section{Conclusion}
\label{Concl} 

In this paper, we propose a method to construct solutions of the cosmological field equations by prescribing a specific functional form of the Hubble parameter capable of inducing different types of future cosmological singularities. This approach provides a particularly straightforward framework for analysing the ECs within general relativity, which in turn allows one to characterise the nature and strength of the resulting singularities in relation to the notion of geodesic incompleteness.

Within this approach, we have explicitly constructed two new models driving the cosmological expansion towards a standard SFS and a novel FTPR (finite-time pseudo-rip) singularity. Both singularities exhibit a similar structural behaviour in that the cosmic time of occurrence, the scale factor, and the (effective) energy density remain finite, while the pressure diverges, i.e. $|p| \to \infty$ as $t \to t_s$. On this basis, one might be tempted to classify both events as SFS \cite{Barrow2004} (or Type II singularities in the sense of Ref. \cite{Nojiri2005}). However, we have argued that such a classification is misleading, since the two scenarios exhibit fundamentally different behaviour when analysed from the perspective of the ECs. While the original SFS was introduced as a scenario in which only the DEC is violated \cite{Barrow2004,Barrow2004a}, the FTPR model (\ref{eq:model2}) displays a phantom-dominated future evolution that leads to the violation of all ECs.

The proposed method can be applied to generate more solutions containing a rich set of exotic singularities in cosmology. Some examples of these singular behaviours that have attracted the attention of many authors in recent years (for a review see Refs. \cite{Nojiri2005,MPD2014CC}), but for our study the most appropriate were a big-rip, a little-rip, a little sibling of the big-rip, a pseudo-rip, and a big-freeze. We have also presented some systematics to study the energy conditions in general relativity which allow to characterise the types of singularities in respect to their strength related to the notion of geodesic incompleteness. In fact, both singularities considered in this work are of a weak type and the evolution of the universe can be extended beyond them. 

We have also discussed the global characteristics of the models at hands by using the method of Penrose diagrams. In particular, we have investigated the structure of cosmological horizons (event, particle, and apparent), as well as the asymptotic structure of singularities. Though only a limited time of evolution of the universe between the standard BB singularity at time $t=0$ to the future exotic singularity at time $t_s$ has been addressed, due to the fact of geodesic completeness of exotics, one may extend them symmetrically into a future region evolving towards another big-crunch singularity as in Ref. \cite{GRG2018}; thus, making kind of a cyclic universe modelling. In addition, we have also found by using the method of the so-called Raychaudhuri averaging that the two models under study are also weak with respect to this definition. 

Finally, we applied this method of generating new weak singularities to construct a physically realistic model containing the standard radiation, dust and cosmological constant components which reproduces the past background expansion history of the $\Lambda$CDM model while evolving towards a future FTPR singularity. We hope both the method of generating the weak singularity solutions and the method of Raychaudhuri averaging may prove successful tool in future cosmological investigations including their observational verification.

\acknowledgements

T.~B.~V. acknowledge the hospitality of the University of Szczecin where this work was partially conducted. He also acknowledges financial support from Universidad Complutense de Madrid and Banco de Santander through Grant No. CT63/19-CT64/19 during that period. We also thank an anonymous Referee for constructive suggestions which led to the improvement of the paper.

\bibliography{SFS_2_2026}{}

\begin{thebibliography}{38}%
\makeatletter
\providecommand \@ifxundefined [1]{%
 \@ifx{#1\undefined}
}%
\providecommand \@ifnum [1]{%
 \ifnum #1\expandafter \@firstoftwo
 \else \expandafter \@secondoftwo
 \fi
}%
\providecommand \@ifx [1]{%
 \ifx #1\expandafter \@firstoftwo
 \else \expandafter \@secondoftwo
 \fi
}%
\providecommand \natexlab [1]{#1}%
\providecommand \enquote  [1]{``#1''}%
\providecommand \bibnamefont  [1]{#1}%
\providecommand \bibfnamefont [1]{#1}%
\providecommand \citenamefont [1]{#1}%
\providecommand \href@noop [0]{\@secondoftwo}%
\providecommand \href [0]{\begingroup \@sanitize@url \@href}%
\providecommand \@href[1]{\@@startlink{#1}\@@href}%
\providecommand \@@href[1]{\endgroup#1\@@endlink}%
\providecommand \@sanitize@url [0]{\catcode `\\12\catcode `\$12\catcode
  `\&12\catcode `\#12\catcode `\^12\catcode `\_12\catcode `\%12\relax}%
\providecommand \@@startlink[1]{}%
\providecommand \@@endlink[0]{}%
\providecommand \url  [0]{\begingroup\@sanitize@url \@url }%
\providecommand \@url [1]{\endgroup\@href {#1}{\urlprefix }}%
\providecommand \urlprefix  [0]{URL }%
\providecommand \Eprint [0]{\href }%
\providecommand \doibase [0]{https://doi.org/}%
\providecommand \selectlanguage [0]{\@gobble}%
\providecommand \bibinfo  [0]{\@secondoftwo}%
\providecommand \bibfield  [0]{\@secondoftwo}%
\providecommand \translation [1]{[#1]}%
\providecommand \BibitemOpen [0]{}%
\providecommand \bibitemStop [0]{}%
\providecommand \bibitemNoStop [0]{.\EOS\space}%
\providecommand \EOS [0]{\spacefactor3000\relax}%
\providecommand \BibitemShut  [1]{\csname bibitem#1\endcsname}%
\let\auto@bib@innerbib\@empty
\bibitem [{\citenamefont {Riess}\ \emph {et~al.}(1998)\citenamefont {Riess},
  \citenamefont {Filippenko}, \citenamefont {Challis}, \citenamefont
  {Clocchiatti}, \citenamefont {Diercks}, \citenamefont {Garnavich},
  \citenamefont {Gilliland}, \citenamefont {Hogan}, \citenamefont {Jha},
  \citenamefont {Kirshner}, \citenamefont {Leibundgut}, \citenamefont
  {Phillips}, \citenamefont {Reiss}, \citenamefont {Schmidt}, \citenamefont
  {Schommer}, \citenamefont {Smith}, \citenamefont {Spyromilio}, \citenamefont
  {Stubbs}, \citenamefont {Suntzeff},\ and\ \citenamefont {Tonry}}]{Riess98}%
  \BibitemOpen
  \bibfield  {author} {\bibinfo {author} {\bibfnamefont {A.~G.}\ \bibnamefont
  {Riess}}, \bibinfo {author} {\bibfnamefont {A.~V.}\ \bibnamefont
  {Filippenko}}, \bibinfo {author} {\bibfnamefont {P.}~\bibnamefont {Challis}},
  \bibinfo {author} {\bibfnamefont {A.}~\bibnamefont {Clocchiatti}}, \bibinfo
  {author} {\bibfnamefont {A.}~\bibnamefont {Diercks}}, \bibinfo {author}
  {\bibfnamefont {P.~M.}\ \bibnamefont {Garnavich}}, \bibinfo {author}
  {\bibfnamefont {R.~L.}\ \bibnamefont {Gilliland}}, \bibinfo {author}
  {\bibfnamefont {C.~J.}\ \bibnamefont {Hogan}}, \bibinfo {author}
  {\bibfnamefont {S.}~\bibnamefont {Jha}}, \bibinfo {author} {\bibfnamefont
  {R.~P.}\ \bibnamefont {Kirshner}}, \bibinfo {author} {\bibfnamefont
  {B.}~\bibnamefont {Leibundgut}}, \bibinfo {author} {\bibfnamefont {M.~M.}\
  \bibnamefont {Phillips}}, \bibinfo {author} {\bibfnamefont {D.}~\bibnamefont
  {Reiss}}, \bibinfo {author} {\bibfnamefont {B.~P.}\ \bibnamefont {Schmidt}},
  \bibinfo {author} {\bibfnamefont {R.~A.}\ \bibnamefont {Schommer}}, \bibinfo
  {author} {\bibfnamefont {R.~C.}\ \bibnamefont {Smith}}, \bibinfo {author}
  {\bibfnamefont {J.}~\bibnamefont {Spyromilio}}, \bibinfo {author}
  {\bibfnamefont {C.}~\bibnamefont {Stubbs}}, \bibinfo {author} {\bibfnamefont
  {N.~B.}\ \bibnamefont {Suntzeff}},\ and\ \bibinfo {author} {\bibfnamefont
  {J.}~\bibnamefont {Tonry}},\ }\bibfield  {title} {\bibinfo {title}
  {Observational evidence from supernovae for an accelerating universe and a
  cosmological constant},\ }\href {https://doi.org/10.1086/300499} {\bibfield
  {journal} {\bibinfo  {journal} {The Astronomical Journal}\ }\textbf {\bibinfo
  {volume} {116}},\ \bibinfo {pages} {1009} (\bibinfo {year}
  {1998})}\BibitemShut {NoStop}%
\bibitem [{\citenamefont {Perlmutter}\ \emph {et~al.}(1999)\citenamefont
  {Perlmutter}, \citenamefont {Aldering}, \citenamefont {Goldhaber},
  \citenamefont {Knop}, \citenamefont {Nugent}, \citenamefont {Castro},
  \citenamefont {Deustua}, \citenamefont {Fabbro}, \citenamefont {Goobar},
  \citenamefont {Groom}, \citenamefont {Hook}, \citenamefont {Kim},
  \citenamefont {Kim}, \citenamefont {Lee}, \citenamefont {Nunes},
  \citenamefont {Pain}, \citenamefont {Pennypacker}, \citenamefont {Quimby},
  \citenamefont {Lidman}, \citenamefont {Ellis}, \citenamefont {Irwin},
  \citenamefont {McMahon}, \citenamefont {Ruiz-Lapuente}, \citenamefont
  {Walton}, \citenamefont {Schaefer}, \citenamefont {Boyle}, \citenamefont
  {Filippenko}, \citenamefont {Matheson}, \citenamefont {Fruchter},
  \citenamefont {Panagia}, \citenamefont {Newberg}, \citenamefont {Couch},\
  and\ \citenamefont {Project}}]{Perlmutter99}%
  \BibitemOpen
  \bibfield  {author} {\bibinfo {author} {\bibfnamefont {S.}~\bibnamefont
  {Perlmutter}}, \bibinfo {author} {\bibfnamefont {G.}~\bibnamefont
  {Aldering}}, \bibinfo {author} {\bibfnamefont {G.}~\bibnamefont {Goldhaber}},
  \bibinfo {author} {\bibfnamefont {R.~A.}\ \bibnamefont {Knop}}, \bibinfo
  {author} {\bibfnamefont {P.}~\bibnamefont {Nugent}}, \bibinfo {author}
  {\bibfnamefont {P.~G.}\ \bibnamefont {Castro}}, \bibinfo {author}
  {\bibfnamefont {S.}~\bibnamefont {Deustua}}, \bibinfo {author} {\bibfnamefont
  {S.}~\bibnamefont {Fabbro}}, \bibinfo {author} {\bibfnamefont
  {A.}~\bibnamefont {Goobar}}, \bibinfo {author} {\bibfnamefont {D.~E.}\
  \bibnamefont {Groom}}, \bibinfo {author} {\bibfnamefont {I.~M.}\ \bibnamefont
  {Hook}}, \bibinfo {author} {\bibfnamefont {A.~G.}\ \bibnamefont {Kim}},
  \bibinfo {author} {\bibfnamefont {M.~Y.}\ \bibnamefont {Kim}}, \bibinfo
  {author} {\bibfnamefont {J.~C.}\ \bibnamefont {Lee}}, \bibinfo {author}
  {\bibfnamefont {N.~J.}\ \bibnamefont {Nunes}}, \bibinfo {author}
  {\bibfnamefont {R.}~\bibnamefont {Pain}}, \bibinfo {author} {\bibfnamefont
  {C.~R.}\ \bibnamefont {Pennypacker}}, \bibinfo {author} {\bibfnamefont
  {R.}~\bibnamefont {Quimby}}, \bibinfo {author} {\bibfnamefont
  {C.}~\bibnamefont {Lidman}}, \bibinfo {author} {\bibfnamefont {R.~S.}\
  \bibnamefont {Ellis}}, \bibinfo {author} {\bibfnamefont {M.}~\bibnamefont
  {Irwin}}, \bibinfo {author} {\bibfnamefont {R.~G.}\ \bibnamefont {McMahon}},
  \bibinfo {author} {\bibfnamefont {P.}~\bibnamefont {Ruiz-Lapuente}}, \bibinfo
  {author} {\bibfnamefont {N.}~\bibnamefont {Walton}}, \bibinfo {author}
  {\bibfnamefont {B.}~\bibnamefont {Schaefer}}, \bibinfo {author}
  {\bibfnamefont {B.~J.}\ \bibnamefont {Boyle}}, \bibinfo {author}
  {\bibfnamefont {A.~V.}\ \bibnamefont {Filippenko}}, \bibinfo {author}
  {\bibfnamefont {T.}~\bibnamefont {Matheson}}, \bibinfo {author}
  {\bibfnamefont {A.~S.}\ \bibnamefont {Fruchter}}, \bibinfo {author}
  {\bibfnamefont {N.}~\bibnamefont {Panagia}}, \bibinfo {author} {\bibfnamefont
  {H.~J.~M.}\ \bibnamefont {Newberg}}, \bibinfo {author} {\bibfnamefont
  {W.~J.}\ \bibnamefont {Couch}},\ and\ \bibinfo {author} {\bibfnamefont
  {T.~S.~C.}\ \bibnamefont {Project}},\ }\bibfield  {title} {\bibinfo {title}
  {Measurements of {$\Omega$} and {$\Lambda$} from 42 high-redshift
  supernovae},\ }\href {https://doi.org/10.1086/307221} {\bibfield  {journal}
  {\bibinfo  {journal} {The Astrophysical Journal}\ }\textbf {\bibinfo {volume}
  {517}},\ \bibinfo {pages} {565} (\bibinfo {year} {1999})}\BibitemShut
  {NoStop}%
\bibitem [{\citenamefont {Di~Valentino}\ \emph {et~al.}(2021)\citenamefont
  {Di~Valentino}, \citenamefont {Mena}, \citenamefont {Pan}, \citenamefont
  {Visinelli}, \citenamefont {Yang}, \citenamefont {Melchiorri}, \citenamefont
  {Mota}, \citenamefont {Riess},\ and\ \citenamefont {Silk}}]{tension}%
  \BibitemOpen
  \bibfield  {author} {\bibinfo {author} {\bibfnamefont {E.}~\bibnamefont
  {Di~Valentino}}, \bibinfo {author} {\bibfnamefont {O.}~\bibnamefont {Mena}},
  \bibinfo {author} {\bibfnamefont {S.}~\bibnamefont {Pan}}, \bibinfo {author}
  {\bibfnamefont {L.}~\bibnamefont {Visinelli}}, \bibinfo {author}
  {\bibfnamefont {W.}~\bibnamefont {Yang}}, \bibinfo {author} {\bibfnamefont
  {A.}~\bibnamefont {Melchiorri}}, \bibinfo {author} {\bibfnamefont {D.~F.}\
  \bibnamefont {Mota}}, \bibinfo {author} {\bibfnamefont {A.~G.}\ \bibnamefont
  {Riess}},\ and\ \bibinfo {author} {\bibfnamefont {J.}~\bibnamefont {Silk}},\
  }\bibfield  {title} {\bibinfo {title} {In the realm of the hubble tension --
  a review of solutions},\ }\href {https://doi.org/10.1088/1361-6382/ac086d}
  {\bibfield  {journal} {\bibinfo  {journal} {Classical and Quantum Gravity}\
  }\textbf {\bibinfo {volume} {38}},\ \bibinfo {pages} {153001} (\bibinfo
  {year} {2021})}\BibitemShut {NoStop}%
\bibitem [{\citenamefont {Weinberg}(1989)}]{CC}%
  \BibitemOpen
  \bibfield  {author} {\bibinfo {author} {\bibfnamefont {S.}~\bibnamefont
  {Weinberg}},\ }\bibfield  {title} {\bibinfo {title} {The cosmological
  constant problem},\ }\href {https://doi.org/10.1103/RevModPhys.61.1}
  {\bibfield  {journal} {\bibinfo  {journal} {Rev. Mod. Phys.}\ }\textbf
  {\bibinfo {volume} {61}},\ \bibinfo {pages} {1} (\bibinfo {year}
  {1989})}\BibitemShut {NoStop}%
\bibitem [{\citenamefont {Adler}\ \emph {et~al.}(1995)\citenamefont {Adler},
  \citenamefont {Casey},\ and\ \citenamefont {Jacob}}]{CC2}%
  \BibitemOpen
  \bibfield  {author} {\bibinfo {author} {\bibfnamefont {R.}~\bibnamefont
  {Adler}}, \bibinfo {author} {\bibfnamefont {B.}~\bibnamefont {Casey}},\ and\
  \bibinfo {author} {\bibfnamefont {O.}~\bibnamefont {Jacob}},\ }\bibfield
  {title} {\bibinfo {title} {Vacuum catastrophe: An elementary exposition of
  the cosmological constant problem},\ }\href@noop {} {\bibfield  {journal}
  {\bibinfo  {journal} {Am. J. Phys.}\ }\textbf {\bibinfo {volume} {63}},\
  \bibinfo {pages} {620} (\bibinfo {year} {1995})}\BibitemShut {NoStop}%
\bibitem [{\citenamefont {Bull}\ \emph {et~al.}(2016)\citenamefont {Bull},
  \citenamefont {Akrami}, \citenamefont {Adamek}, \citenamefont {Baker},
  \citenamefont {Bellini}, \citenamefont {{Beltrán Jiménez}}, \citenamefont
  {Bentivegna}, \citenamefont {Camera}, \citenamefont {Clesse}, \citenamefont
  {Davis}, \citenamefont {{Di Dio}}, \citenamefont {Enander}, \citenamefont
  {Heavens}, \citenamefont {Heisenberg}, \citenamefont {Hu}, \citenamefont
  {Llinares}, \citenamefont {Maartens}, \citenamefont {Mörtsell},
  \citenamefont {Nadathur}, \citenamefont {Noller}, \citenamefont {Pasechnik},
  \citenamefont {Pawlowski}, \citenamefont {Pereira}, \citenamefont {Quartin},
  \citenamefont {Ricciardone}, \citenamefont {Riemer-Sørensen}, \citenamefont
  {Rinaldi}, \citenamefont {Sakstein}, \citenamefont {Saltas}, \citenamefont
  {Salzano}, \citenamefont {Sawicki}, \citenamefont {Solomon}, \citenamefont
  {Spolyar}, \citenamefont {Starkman}, \citenamefont {Steer}, \citenamefont
  {Tereno}, \citenamefont {Verde}, \citenamefont {Villaescusa-Navarro},
  \citenamefont {{von Strauss}},\ and\ \citenamefont {Winther}}]{+LCDM}%
  \BibitemOpen
  \bibfield  {author} {\bibinfo {author} {\bibfnamefont {P.}~\bibnamefont
  {Bull}}, \bibinfo {author} {\bibfnamefont {Y.}~\bibnamefont {Akrami}},
  \bibinfo {author} {\bibfnamefont {J.}~\bibnamefont {Adamek}}, \bibinfo
  {author} {\bibfnamefont {T.}~\bibnamefont {Baker}}, \bibinfo {author}
  {\bibfnamefont {E.}~\bibnamefont {Bellini}}, \bibinfo {author} {\bibfnamefont
  {J.}~\bibnamefont {{Beltrán Jiménez}}}, \bibinfo {author} {\bibfnamefont
  {E.}~\bibnamefont {Bentivegna}}, \bibinfo {author} {\bibfnamefont
  {S.}~\bibnamefont {Camera}}, \bibinfo {author} {\bibfnamefont
  {S.}~\bibnamefont {Clesse}}, \bibinfo {author} {\bibfnamefont {J.~H.}\
  \bibnamefont {Davis}}, \bibinfo {author} {\bibfnamefont {E.}~\bibnamefont
  {{Di Dio}}}, \bibinfo {author} {\bibfnamefont {J.}~\bibnamefont {Enander}},
  \bibinfo {author} {\bibfnamefont {A.}~\bibnamefont {Heavens}}, \bibinfo
  {author} {\bibfnamefont {L.}~\bibnamefont {Heisenberg}}, \bibinfo {author}
  {\bibfnamefont {B.}~\bibnamefont {Hu}}, \bibinfo {author} {\bibfnamefont
  {C.}~\bibnamefont {Llinares}}, \bibinfo {author} {\bibfnamefont
  {R.}~\bibnamefont {Maartens}}, \bibinfo {author} {\bibfnamefont
  {E.}~\bibnamefont {Mörtsell}}, \bibinfo {author} {\bibfnamefont
  {S.}~\bibnamefont {Nadathur}}, \bibinfo {author} {\bibfnamefont
  {J.}~\bibnamefont {Noller}}, \bibinfo {author} {\bibfnamefont
  {R.}~\bibnamefont {Pasechnik}}, \bibinfo {author} {\bibfnamefont {M.~S.}\
  \bibnamefont {Pawlowski}}, \bibinfo {author} {\bibfnamefont {T.~S.}\
  \bibnamefont {Pereira}}, \bibinfo {author} {\bibfnamefont {M.}~\bibnamefont
  {Quartin}}, \bibinfo {author} {\bibfnamefont {A.}~\bibnamefont
  {Ricciardone}}, \bibinfo {author} {\bibfnamefont {S.}~\bibnamefont
  {Riemer-Sørensen}}, \bibinfo {author} {\bibfnamefont {M.}~\bibnamefont
  {Rinaldi}}, \bibinfo {author} {\bibfnamefont {J.}~\bibnamefont {Sakstein}},
  \bibinfo {author} {\bibfnamefont {I.~D.}\ \bibnamefont {Saltas}}, \bibinfo
  {author} {\bibfnamefont {V.}~\bibnamefont {Salzano}}, \bibinfo {author}
  {\bibfnamefont {I.}~\bibnamefont {Sawicki}}, \bibinfo {author} {\bibfnamefont
  {A.~R.}\ \bibnamefont {Solomon}}, \bibinfo {author} {\bibfnamefont
  {D.}~\bibnamefont {Spolyar}}, \bibinfo {author} {\bibfnamefont {G.~D.}\
  \bibnamefont {Starkman}}, \bibinfo {author} {\bibfnamefont {D.}~\bibnamefont
  {Steer}}, \bibinfo {author} {\bibfnamefont {I.}~\bibnamefont {Tereno}},
  \bibinfo {author} {\bibfnamefont {L.}~\bibnamefont {Verde}}, \bibinfo
  {author} {\bibfnamefont {F.}~\bibnamefont {Villaescusa-Navarro}}, \bibinfo
  {author} {\bibfnamefont {M.}~\bibnamefont {{von Strauss}}},\ and\ \bibinfo
  {author} {\bibfnamefont {H.~A.}\ \bibnamefont {Winther}},\ }\bibfield
  {title} {\bibinfo {title} {Beyond {$\Lambda$CDM}: Problems, solutions, and
  the road ahead},\ }\href
  {https://doi.org/https://doi.org/10.1016/j.dark.2016.02.001} {\bibfield
  {journal} {\bibinfo  {journal} {Physics of the Dark Universe}\ }\textbf
  {\bibinfo {volume} {12}},\ \bibinfo {pages} {56} (\bibinfo {year}
  {2016})}\BibitemShut {NoStop}%
\bibitem [{\citenamefont {Perivolaropoulos}\ and\ \citenamefont
  {Skara}(2022)}]{+LCDM2}%
  \BibitemOpen
  \bibfield  {author} {\bibinfo {author} {\bibfnamefont {L.}~\bibnamefont
  {Perivolaropoulos}}\ and\ \bibinfo {author} {\bibfnamefont {F.}~\bibnamefont
  {Skara}},\ }\bibfield  {title} {\bibinfo {title} {Challenges for
  {$\Lambda$}cdm: An update},\ }\href
  {https://doi.org/https://doi.org/10.1016/j.newar.2022.101659} {\bibfield
  {journal} {\bibinfo  {journal} {New Astronomy Reviews}\ }\textbf {\bibinfo
  {volume} {95}},\ \bibinfo {pages} {101659} (\bibinfo {year}
  {2022})}\BibitemShut {NoStop}%
\bibitem [{\citenamefont {Caldwell}(2002)}]{Caldwell2002}%
  \BibitemOpen
  \bibfield  {author} {\bibinfo {author} {\bibfnamefont {R.}~\bibnamefont
  {Caldwell}},\ }\bibfield  {title} {\bibinfo {title} {A phantom menace?
  cosmological consequences of a dark energy component with super-negative
  equation of state},\ }\href
  {https://doi.org/http://dx.doi.org/10.1016/S0370-2693(02)02589-3} {\bibfield
  {journal} {\bibinfo  {journal} {Physics Letters B}\ }\textbf {\bibinfo
  {volume} {545}},\ \bibinfo {pages} {23 } (\bibinfo {year}
  {2002})}\BibitemShut {NoStop}%
\bibitem [{\citenamefont {Caldwell}\ \emph {et~al.}(2003)\citenamefont
  {Caldwell}, \citenamefont {Kamionkowski},\ and\ \citenamefont
  {Weinberg}}]{Caldwell2003}%
  \BibitemOpen
  \bibfield  {author} {\bibinfo {author} {\bibfnamefont {R.~R.}\ \bibnamefont
  {Caldwell}}, \bibinfo {author} {\bibfnamefont {M.}~\bibnamefont
  {Kamionkowski}},\ and\ \bibinfo {author} {\bibfnamefont {N.~N.}\ \bibnamefont
  {Weinberg}},\ }\bibfield  {title} {\bibinfo {title} {Phantom energy: Dark
  energy with $w=-1$ causes a cosmic doomsday},\ }\href
  {https://doi.org/10.1103/PhysRevLett.91.071301} {\bibfield  {journal}
  {\bibinfo  {journal} {Phys. Rev. Lett.}\ }\textbf {\bibinfo {volume} {91}},\
  \bibinfo {pages} {071301} (\bibinfo {year} {2003})}\BibitemShut {NoStop}%
\bibitem [{\citenamefont {D\c{a}browski}\ \emph {et~al.}(2003)\citenamefont
  {D\c{a}browski}, \citenamefont {Stachowiak},\ and\ \citenamefont
  {Szyd\l{}owski}}]{MPD2003}%
  \BibitemOpen
  \bibfield  {author} {\bibinfo {author} {\bibfnamefont {M.~P.}\ \bibnamefont
  {D\c{a}browski}}, \bibinfo {author} {\bibfnamefont {T.}~\bibnamefont
  {Stachowiak}},\ and\ \bibinfo {author} {\bibfnamefont {M.}~\bibnamefont
  {Szyd\l{}owski}},\ }\bibfield  {title} {\bibinfo {title} {Phantom
  cosmologies},\ }\href {https://doi.org/10.1103/PhysRevD.68.103519} {\bibfield
   {journal} {\bibinfo  {journal} {Phys. Rev. D}\ }\textbf {\bibinfo {volume}
  {68}},\ \bibinfo {pages} {103519} (\bibinfo {year} {2003})}\BibitemShut
  {NoStop}%
\bibitem [{\citenamefont {Hawking}\ and\ \citenamefont {Ellis}(1973)}]{HE}%
  \BibitemOpen
  \bibfield  {author} {\bibinfo {author} {\bibfnamefont {S.~W.}\ \bibnamefont
  {Hawking}}\ and\ \bibinfo {author} {\bibfnamefont {G.~F.~R.}\ \bibnamefont
  {Ellis}},\ }\href {http://dx.doi.org/10.1017/CBO9780511524646} {\emph
  {\bibinfo {title} {The Large Scale Structure of Space-Time}}}\ (\bibinfo
  {publisher} {Cambridge University Press},\ \bibinfo {year} {1973})\ \bibinfo
  {note} {cambridge Books Online}\BibitemShut {NoStop}%
\bibitem [{\citenamefont {Barrow}(2004{\natexlab{a}})}]{Barrow2004}%
  \BibitemOpen
  \bibfield  {author} {\bibinfo {author} {\bibfnamefont {J.~D.}\ \bibnamefont
  {Barrow}},\ }\bibfield  {title} {\bibinfo {title} {Sudden future
  singularities},\ }\href {http://stacks.iop.org/0264-9381/21/i=11/a=L03}
  {\bibfield  {journal} {\bibinfo  {journal} {Classical and Quantum Gravity}\
  }\textbf {\bibinfo {volume} {21}},\ \bibinfo {pages} {L79} (\bibinfo {year}
  {2004}{\natexlab{a}})}\BibitemShut {NoStop}%
\bibitem [{\citenamefont {Barrow}(2004{\natexlab{b}})}]{Barrow2004a}%
  \BibitemOpen
  \bibfield  {author} {\bibinfo {author} {\bibfnamefont {J.~D.}\ \bibnamefont
  {Barrow}},\ }\bibfield  {title} {\bibinfo {title} {More general sudden
  singularities},\ }\href {https://doi.org/10.1088/0264-9381/21/23/020}
  {\bibfield  {journal} {\bibinfo  {journal} {Classical and Quantum Gravity}\
  }\textbf {\bibinfo {volume} {21}},\ \bibinfo {pages} {5619} (\bibinfo {year}
  {2004}{\natexlab{b}})}\BibitemShut {NoStop}%
\bibitem [{\citenamefont {Nojiri}\ \emph {et~al.}(2005)\citenamefont {Nojiri},
  \citenamefont {Odintsov},\ and\ \citenamefont {Tsujikawa}}]{Nojiri2005}%
  \BibitemOpen
  \bibfield  {author} {\bibinfo {author} {\bibfnamefont {S.}~\bibnamefont
  {Nojiri}}, \bibinfo {author} {\bibfnamefont {S.~D.}\ \bibnamefont
  {Odintsov}},\ and\ \bibinfo {author} {\bibfnamefont {S.}~\bibnamefont
  {Tsujikawa}},\ }\bibfield  {title} {\bibinfo {title} {Properties of
  singularities in the (phantom) dark energy universe},\ }\href
  {https://doi.org/10.1103/PhysRevD.71.063004} {\bibfield  {journal} {\bibinfo
  {journal} {Phys. Rev. D}\ }\textbf {\bibinfo {volume} {71}},\ \bibinfo
  {pages} {063004} (\bibinfo {year} {2005})}\BibitemShut {NoStop}%
\bibitem [{\citenamefont {D\c{a}browski}(2014)}]{MPD2014CC}%
  \BibitemOpen
  \bibfield  {author} {\bibinfo {author} {\bibfnamefont {M.}~\bibnamefont
  {D\c{a}browski}},\ }\bibfield  {title} {\bibinfo {title} {Are singularities
  the limits of cosmology?},\ }in\ \href@noop {} {\emph {\bibinfo {booktitle}
  {Mathematical Structures of the Universe}}},\ \bibinfo {editor} {edited by\
  \bibinfo {editor} {\bibfnamefont {M.}~\bibnamefont {Heller}}, \bibinfo
  {editor} {\bibfnamefont {M.}~\bibnamefont {Eckstein}},\ and\ \bibinfo
  {editor} {\bibfnamefont {S.}~\bibnamefont {Szybka}}}\ (\bibinfo  {publisher}
  {Copernicus Center Press, Krak\'ow},\ \bibinfo {year} {2014})\ pp.\ \bibinfo
  {pages} {101--118}\BibitemShut {NoStop}%
\bibitem [{\citenamefont {Frampton}\ \emph {et~al.}(2012)\citenamefont
  {Frampton}, \citenamefont {Ludwick},\ and\ \citenamefont {Scherrer}}]{PR}%
  \BibitemOpen
  \bibfield  {author} {\bibinfo {author} {\bibfnamefont {P.~H.}\ \bibnamefont
  {Frampton}}, \bibinfo {author} {\bibfnamefont {K.~J.}\ \bibnamefont
  {Ludwick}},\ and\ \bibinfo {author} {\bibfnamefont {R.~J.}\ \bibnamefont
  {Scherrer}},\ }\bibfield  {title} {\bibinfo {title} {Pseudo-rip: Cosmological
  models intermediate between the cosmological constant and the little rip},\
  }\href {https://doi.org/10.1103/PhysRevD.85.083001} {\bibfield  {journal}
  {\bibinfo  {journal} {Phys. Rev. D}\ }\textbf {\bibinfo {volume} {85}},\
  \bibinfo {pages} {083001} (\bibinfo {year} {2012})}\BibitemShut {NoStop}%
\bibitem [{\citenamefont {Trivedi}\ and\ \citenamefont
  {Timoshkin}(2024)}]{PR1}%
  \BibitemOpen
  \bibfield  {author} {\bibinfo {author} {\bibfnamefont {O.}~\bibnamefont
  {Trivedi}}\ and\ \bibinfo {author} {\bibfnamefont {A.~V.}\ \bibnamefont
  {Timoshkin}},\ }\bibfield  {title} {\bibinfo {title} {Little rip, pseudo rip
  and bounce cosmology with generalized equation of state in non-standard
  backgrounds},\ }\href {https://doi.org/10.1140/epjc/s10052-024-12640-w}
  {\bibfield  {journal} {\bibinfo  {journal} {Eur. Phys. J. C}\ }\textbf
  {\bibinfo {volume} {84}},\ \bibinfo {pages} {277} (\bibinfo {year}
  {2024})}\BibitemShut {NoStop}%
\bibitem [{\citenamefont {Bouhmadi-L\'{o}pez}\ \emph
  {et~al.}(2015)\citenamefont {Bouhmadi-L\'{o}pez}, \citenamefont {Errahmani},
  \citenamefont {Mart\'{\i}n-Moruno}, \citenamefont {Ouali},\ and\
  \citenamefont {Tavakoli}}]{LSBR}%
  \BibitemOpen
  \bibfield  {author} {\bibinfo {author} {\bibfnamefont {M.}~\bibnamefont
  {Bouhmadi-L\'{o}pez}}, \bibinfo {author} {\bibfnamefont {A.}~\bibnamefont
  {Errahmani}}, \bibinfo {author} {\bibfnamefont {P.}~\bibnamefont
  {Mart\'{\i}n-Moruno}}, \bibinfo {author} {\bibfnamefont {T.}~\bibnamefont
  {Ouali}},\ and\ \bibinfo {author} {\bibfnamefont {Y.}~\bibnamefont
  {Tavakoli}},\ }\bibfield  {title} {\bibinfo {title} {The little sibling of
  the big rip singularity},\ }\href {https://doi.org/10.1142/S0218271815500789}
  {\bibfield  {journal} {\bibinfo  {journal} {International Journal of Modern
  Physics D}\ }\textbf {\bibinfo {volume} {24}},\ \bibinfo {pages} {1550078}
  (\bibinfo {year} {2015})},\ \Eprint
  {https://arxiv.org/abs/https://doi.org/10.1142/S0218271815500789}
  {https://doi.org/10.1142/S0218271815500789} \BibitemShut {NoStop}%
\bibitem [{\citenamefont {Frampton}\ \emph {et~al.}(2011)\citenamefont
  {Frampton}, \citenamefont {Ludwick},\ and\ \citenamefont {Scherrer}}]{LR}%
  \BibitemOpen
  \bibfield  {author} {\bibinfo {author} {\bibfnamefont {P.~H.}\ \bibnamefont
  {Frampton}}, \bibinfo {author} {\bibfnamefont {K.~J.}\ \bibnamefont
  {Ludwick}},\ and\ \bibinfo {author} {\bibfnamefont {R.~J.}\ \bibnamefont
  {Scherrer}},\ }\bibfield  {title} {\bibinfo {title} {The little rip},\ }\href
  {https://doi.org/10.1103/PhysRevD.84.063003} {\bibfield  {journal} {\bibinfo
  {journal} {Phys. Rev. D}\ }\textbf {\bibinfo {volume} {84}},\ \bibinfo
  {pages} {063003} (\bibinfo {year} {2011})}\BibitemShut {NoStop}%
\bibitem [{\citenamefont {Bouhmadi-L\'opez}\ \emph {et~al.}(2008)\citenamefont
  {Bouhmadi-L\'opez}, \citenamefont {Gonz\'alez-D\'iaz},\ and\ \citenamefont
  {Martín-Moruno}}]{BF}%
  \BibitemOpen
  \bibfield  {author} {\bibinfo {author} {\bibfnamefont {M.}~\bibnamefont
  {Bouhmadi-L\'opez}}, \bibinfo {author} {\bibfnamefont {P.}~\bibnamefont
  {Gonz\'alez-D\'iaz}},\ and\ \bibinfo {author} {\bibfnamefont
  {P.}~\bibnamefont {Martín-Moruno}},\ }\bibfield  {title} {\bibinfo {title}
  {Worse than a big rip?},\ }\href
  {https://doi.org/https://doi.org/10.1016/j.physletb.2007.10.079} {\bibfield
  {journal} {\bibinfo  {journal} {Physics Letters B}\ }\textbf {\bibinfo
  {volume} {659}},\ \bibinfo {pages} {1} (\bibinfo {year} {2008})}\BibitemShut
  {NoStop}%
\bibitem [{\citenamefont {Bouhmadi-L\'{o}pez}\ \emph
  {et~al.}(2008)\citenamefont {Bouhmadi-L\'{o}pez}, \citenamefont
  {Gonz\'{a}lez-D\'{i}az},\ and\ \citenamefont {Mart\'{i}n-Moruno}}]{BF1}%
  \BibitemOpen
  \bibfield  {author} {\bibinfo {author} {\bibfnamefont {M.}~\bibnamefont
  {Bouhmadi-L\'{o}pez}}, \bibinfo {author} {\bibfnamefont {P.~F.}\ \bibnamefont
  {Gonz\'{a}lez-D\'{i}az}},\ and\ \bibinfo {author} {\bibfnamefont
  {P.}~\bibnamefont {Mart\'{i}n-Moruno}},\ }\bibfield  {title} {\bibinfo
  {title} {On the generalized chaplygin gas: worse than a big-rip or quiaeter
  than a sudden singularity?},\ }\href
  {https://doi.org/10.1142/S0218271808013856} {\bibfield  {journal} {\bibinfo
  {journal} {International Journal of Modern Physics D}\ }\textbf {\bibinfo
  {volume} {17}},\ \bibinfo {pages} {2269} (\bibinfo {year}
  {2008})}\BibitemShut {NoStop}%
\bibitem [{\citenamefont {Gorini}\ \emph {et~al.}(2004)\citenamefont {Gorini},
  \citenamefont {Kamenshchik}, \citenamefont {Moschella},\ and\ \citenamefont
  {Pasquier}}]{big-brake}%
  \BibitemOpen
  \bibfield  {author} {\bibinfo {author} {\bibfnamefont {V.}~\bibnamefont
  {Gorini}}, \bibinfo {author} {\bibfnamefont {A.}~\bibnamefont {Kamenshchik}},
  \bibinfo {author} {\bibfnamefont {U.}~\bibnamefont {Moschella}},\ and\
  \bibinfo {author} {\bibfnamefont {V.}~\bibnamefont {Pasquier}},\ }\bibfield
  {title} {\bibinfo {title} {Tachyons, scalar fields, and cosmology},\ }\href
  {https://doi.org/10.1103/PhysRevD.69.123512} {\bibfield  {journal} {\bibinfo
  {journal} {Phys. Rev. D}\ }\textbf {\bibinfo {volume} {69}},\ \bibinfo
  {pages} {123512} (\bibinfo {year} {2004})}\BibitemShut {NoStop}%
\bibitem [{\citenamefont {D\c{a}browski}\ \emph {et~al.}(2007)\citenamefont
  {D\c{a}browski}, \citenamefont {Denkiewicz},\ and\ \citenamefont
  {Hendry}}]{Hendry2007}%
  \BibitemOpen
  \bibfield  {author} {\bibinfo {author} {\bibfnamefont {M.~P.}\ \bibnamefont
  {D\c{a}browski}}, \bibinfo {author} {\bibfnamefont {T.}~\bibnamefont
  {Denkiewicz}},\ and\ \bibinfo {author} {\bibfnamefont {M.~A.}\ \bibnamefont
  {Hendry}},\ }\bibfield  {title} {\bibinfo {title} {How far is it to a sudden
  future singularity of pressure?},\ }\href
  {https://doi.org/10.1103/PhysRevD.75.123524} {\bibfield  {journal} {\bibinfo
  {journal} {Phys. Rev. D}\ }\textbf {\bibinfo {volume} {75}},\ \bibinfo
  {pages} {123524} (\bibinfo {year} {2007})}\BibitemShut {NoStop}%
\bibitem [{\citenamefont {Keresztes}\ \emph {et~al.}(2009)\citenamefont
  {Keresztes}, \citenamefont {Gergely}, \citenamefont {Gorini}, \citenamefont
  {Moschella},\ and\ \citenamefont {Kamenshchik}}]{Keresztes2009}%
  \BibitemOpen
  \bibfield  {author} {\bibinfo {author} {\bibfnamefont {Z.}~\bibnamefont
  {Keresztes}}, \bibinfo {author} {\bibfnamefont {L.}~\bibnamefont {Gergely}},
  \bibinfo {author} {\bibfnamefont {V.}~\bibnamefont {Gorini}}, \bibinfo
  {author} {\bibfnamefont {U.}~\bibnamefont {Moschella}},\ and\ \bibinfo
  {author} {\bibfnamefont {A.~Y.}\ \bibnamefont {Kamenshchik}},\ }\bibfield
  {title} {\bibinfo {title} {Tachyon cosmology, supernovae data, and the big
  brake singularity},\ }\href {https://doi.org/10.1103/PhysRevD.79.083504}
  {\bibfield  {journal} {\bibinfo  {journal} {Phys. Rev. D}\ }\textbf {\bibinfo
  {volume} {79}},\ \bibinfo {pages} {083504} (\bibinfo {year}
  {2009})}\BibitemShut {NoStop}%
\bibitem [{\citenamefont {Keresztes}\ \emph {et~al.}(2010)\citenamefont
  {Keresztes}, \citenamefont {Gergely}, \citenamefont {Kamenshchik},
  \citenamefont {Gorini},\ and\ \citenamefont {Polarski}}]{Keresztes2010}%
  \BibitemOpen
  \bibfield  {author} {\bibinfo {author} {\bibfnamefont {Z.}~\bibnamefont
  {Keresztes}}, \bibinfo {author} {\bibfnamefont {L.~A.}\ \bibnamefont
  {Gergely}}, \bibinfo {author} {\bibfnamefont {A.~Y.}\ \bibnamefont
  {Kamenshchik}}, \bibinfo {author} {\bibfnamefont {V.}~\bibnamefont
  {Gorini}},\ and\ \bibinfo {author} {\bibfnamefont {D.}~\bibnamefont
  {Polarski}},\ }\bibfield  {title} {\bibinfo {title} {Will the tachyonic
  universe survive the big brake?},\ }\href
  {https://doi.org/10.1103/PhysRevD.82.123534} {\bibfield  {journal} {\bibinfo
  {journal} {Phys. Rev. D}\ }\textbf {\bibinfo {volume} {82}},\ \bibinfo
  {pages} {123534} (\bibinfo {year} {2010})}\BibitemShut {NoStop}%
\bibitem [{\citenamefont {D\c{a}browski}\ and\ \citenamefont
  {Marosek}(2018)}]{GRG2018}%
  \BibitemOpen
  \bibfield  {author} {\bibinfo {author} {\bibfnamefont {M.}~\bibnamefont
  {D\c{a}browski}}\ and\ \bibinfo {author} {\bibfnamefont {K.}~\bibnamefont
  {Marosek}},\ }\bibfield  {title} {\bibinfo {title} {Non-exotic conformal
  structure of weak exotic singularities},\ }\href
  {https://doi.org/10.1007/s10714-018-2482-1} {\bibfield  {journal} {\bibinfo
  {journal} {General Relativity and Gravitation}\ }\textbf {\bibinfo {volume}
  {50}},\ \bibinfo {pages} {160} (\bibinfo {year} {2018})}\BibitemShut
  {NoStop}%
\bibitem [{\citenamefont {Abramowitz}\ and\ \citenamefont
  {Stegun}(1974)}]{MathFuncBook}%
  \BibitemOpen
  \bibfield  {author} {\bibinfo {author} {\bibfnamefont {M.}~\bibnamefont
  {Abramowitz}}\ and\ \bibinfo {author} {\bibfnamefont {I.~A.}\ \bibnamefont
  {Stegun}},\ }\href@noop {} {\emph {\bibinfo {title} {Handbook of Mathematical
  Functions}}}\ (\bibinfo  {publisher} {Dover Publications, Inc.},\ \bibinfo
  {address} {New York, USA},\ \bibinfo {year} {1974})\BibitemShut {NoStop}%
\bibitem [{\citenamefont {Kanai}\ \emph {et~al.}(2023)\citenamefont {Kanai},
  \citenamefont {Nomura},\ and\ \citenamefont {Yoshida}}]{Yoshida2023}%
  \BibitemOpen
  \bibfield  {author} {\bibinfo {author} {\bibfnamefont {T.}~\bibnamefont
  {Kanai}}, \bibinfo {author} {\bibfnamefont {K.}~\bibnamefont {Nomura}},\ and\
  \bibinfo {author} {\bibfnamefont {D.}~\bibnamefont {Yoshida}},\ }\bibfield
  {title} {\bibinfo {title} {Entropy bound and a geometrically nonsingular
  universe},\ }\href {https://doi.org/10.1103/PhysRevD.108.104024} {\bibfield
  {journal} {\bibinfo  {journal} {Phys. Rev. D}\ }\textbf {\bibinfo {volume}
  {108}},\ \bibinfo {pages} {104024} (\bibinfo {year} {2023})}\BibitemShut
  {NoStop}%
\bibitem [{\citenamefont {Melia}(2018)}]{Melia:2018xtf}%
  \BibitemOpen
  \bibfield  {author} {\bibinfo {author} {\bibfnamefont {F.}~\bibnamefont
  {Melia}},\ }\bibfield  {title} {\bibinfo {title} {The apparent
  (gravitational) horizon in cosmology},\ }\href
  {https://doi.org/10.1119/1.5045333} {\bibfield  {journal} {\bibinfo
  {journal} {American Journal of Physics}\ }\textbf {\bibinfo {volume} {86}},\
  \bibinfo {pages} {585} (\bibinfo {year} {2018})}\BibitemShut {NoStop}%
\bibitem [{\citenamefont {Khoury}\ \emph {et~al.}(2001)\citenamefont {Khoury},
  \citenamefont {Ovrut}, \citenamefont {Steinhardt},\ and\ \citenamefont
  {Turok}}]{khoury2001}%
  \BibitemOpen
  \bibfield  {author} {\bibinfo {author} {\bibfnamefont {J.}~\bibnamefont
  {Khoury}}, \bibinfo {author} {\bibfnamefont {B.~A.}\ \bibnamefont {Ovrut}},
  \bibinfo {author} {\bibfnamefont {P.~J.}\ \bibnamefont {Steinhardt}},\ and\
  \bibinfo {author} {\bibfnamefont {N.}~\bibnamefont {Turok}},\ }\bibfield
  {title} {\bibinfo {title} {Ekpyrotic universe: Colliding branes and the
  origin of the hot big bang},\ }\href
  {https://doi.org/10.1103/PhysRevD.64.123522} {\bibfield  {journal} {\bibinfo
  {journal} {Phys. Rev. D}\ }\textbf {\bibinfo {volume} {64}},\ \bibinfo
  {pages} {123522} (\bibinfo {year} {2001})}\BibitemShut {NoStop}%
\bibitem [{\citenamefont {Steinhardt}\ and\ \citenamefont
  {Turok}(2002)}]{turok2002}%
  \BibitemOpen
  \bibfield  {author} {\bibinfo {author} {\bibfnamefont {P.~J.}\ \bibnamefont
  {Steinhardt}}\ and\ \bibinfo {author} {\bibfnamefont {N.}~\bibnamefont
  {Turok}},\ }\bibfield  {title} {\bibinfo {title} {Cosmic evolution in a
  cyclic universe},\ }\href {https://doi.org/10.1103/PhysRevD.65.126003}
  {\bibfield  {journal} {\bibinfo  {journal} {Phys. Rev. D}\ }\textbf {\bibinfo
  {volume} {65}},\ \bibinfo {pages} {126003} (\bibinfo {year}
  {2002})}\BibitemShut {NoStop}%
\bibitem [{\citenamefont {Khoury}\ \emph {et~al.}(2002)\citenamefont {Khoury},
  \citenamefont {Ovrut}, \citenamefont {Seiberg}, \citenamefont {Steinhardt},\
  and\ \citenamefont {Turok}}]{khoury2002}%
  \BibitemOpen
  \bibfield  {author} {\bibinfo {author} {\bibfnamefont {J.}~\bibnamefont
  {Khoury}}, \bibinfo {author} {\bibfnamefont {B.~A.}\ \bibnamefont {Ovrut}},
  \bibinfo {author} {\bibfnamefont {N.}~\bibnamefont {Seiberg}}, \bibinfo
  {author} {\bibfnamefont {P.~J.}\ \bibnamefont {Steinhardt}},\ and\ \bibinfo
  {author} {\bibfnamefont {N.}~\bibnamefont {Turok}},\ }\bibfield  {title}
  {\bibinfo {title} {From big crunch to big bang},\ }\href
  {https://doi.org/10.1103/PhysRevD.65.086007} {\bibfield  {journal} {\bibinfo
  {journal} {Phys. Rev. D}\ }\textbf {\bibinfo {volume} {65}},\ \bibinfo
  {pages} {086007} (\bibinfo {year} {2002})}\BibitemShut {NoStop}%
\bibitem [{\citenamefont {Fern\'andez-Jambrina}\ and\ \citenamefont
  {Lazkoz}(2004)}]{Leonardo}%
  \BibitemOpen
  \bibfield  {author} {\bibinfo {author} {\bibfnamefont {L.}~\bibnamefont
  {Fern\'andez-Jambrina}}\ and\ \bibinfo {author} {\bibfnamefont
  {R.}~\bibnamefont {Lazkoz}},\ }\bibfield  {title} {\bibinfo {title} {Geodesic
  behavior of sudden future singularities},\ }\href
  {https://doi.org/10.1103/PhysRevD.70.121503} {\bibfield  {journal} {\bibinfo
  {journal} {Phys. Rev. D}\ }\textbf {\bibinfo {volume} {70}},\ \bibinfo
  {pages} {121503} (\bibinfo {year} {2004})}\BibitemShut {NoStop}%
\bibitem [{\citenamefont {Tipler}(1977)}]{Tipler}%
  \BibitemOpen
  \bibfield  {author} {\bibinfo {author} {\bibfnamefont {F.~J.}\ \bibnamefont
  {Tipler}},\ }\bibfield  {title} {\bibinfo {title} {Singularities in
  conformally flat spacetimes},\ }\href
  {https://doi.org/http://dx.doi.org/10.1016/0375-9601(77)90508-4} {\bibfield
  {journal} {\bibinfo  {journal} {Physics Letters A}\ }\textbf {\bibinfo
  {volume} {64}},\ \bibinfo {pages} {8 } (\bibinfo {year} {1977})}\BibitemShut
  {NoStop}%
\bibitem [{\citenamefont {Kr\'olak}(1986)}]{Krolak}%
  \BibitemOpen
  \bibfield  {author} {\bibinfo {author} {\bibfnamefont {A.}~\bibnamefont
  {Kr\'olak}},\ }\bibfield  {title} {\bibinfo {title} {Towards the proof of the
  cosmic censorship hypothesis},\ }\href
  {http://stacks.iop.org/0264-9381/3/i=3/a=004} {\bibfield  {journal} {\bibinfo
   {journal} {Classical and Quantum Gravity}\ }\textbf {\bibinfo {volume}
  {3}},\ \bibinfo {pages} {267} (\bibinfo {year} {1986})}\BibitemShut {NoStop}%
\bibitem [{\citenamefont {Raychaudhuri}(1998)}]{raych98}%
  \BibitemOpen
  \bibfield  {author} {\bibinfo {author} {\bibfnamefont {A.~K.}\ \bibnamefont
  {Raychaudhuri}},\ }\bibfield  {title} {\bibinfo {title} {Theorem for
  nonrotating singularity-free universes},\ }\href
  {https://doi.org/10.1103/PhysRevLett.80.654} {\bibfield  {journal} {\bibinfo
  {journal} {Phys. Rev. Lett.}\ }\textbf {\bibinfo {volume} {80}},\ \bibinfo
  {pages} {654} (\bibinfo {year} {1998})}\BibitemShut {NoStop}%
\bibitem [{\citenamefont {D\c{a}browski}(2011)}]{PLB2011}%
  \BibitemOpen
  \bibfield  {author} {\bibinfo {author} {\bibfnamefont {M.~P.}\ \bibnamefont
  {D\c{a}browski}},\ }\bibfield  {title} {\bibinfo {title} {Spacetime averaging
  of exotic singularity universes},\ }\href
  {https://doi.org/http://dx.doi.org/10.1016/j.physletb.2011.07.043} {\bibfield
   {journal} {\bibinfo  {journal} {Physics Letters B}\ }\textbf {\bibinfo
  {volume} {702}},\ \bibinfo {pages} {320 } (\bibinfo {year}
  {2011})}\BibitemShut {NoStop}%
\bibitem [{\citenamefont {Aghanim}\ \emph {et~al.}(2020)\citenamefont {Aghanim}
  \emph {et~al.}}]{Planck:2018vyg}%
  \BibitemOpen
  \bibfield  {author} {\bibinfo {author} {\bibfnamefont {N.}~\bibnamefont
  {Aghanim}} \emph {et~al.} (\bibinfo {collaboration} {Planck}),\ }\bibfield
  {title} {\bibinfo {title} {{Planck 2018 results. VI. Cosmological
  parameters}},\ }\href {https://doi.org/10.1051/0004-6361/201833910}
  {\bibfield  {journal} {\bibinfo  {journal} {Astron. Astrophys.}\ }\textbf
  {\bibinfo {volume} {641}},\ \bibinfo {pages} {A6} (\bibinfo {year} {2020})},\
  \bibinfo {note} {[Erratum: Astron.Astrophys. 652, C4 (2021)]},\ \Eprint
  {https://arxiv.org/abs/1807.06209} {arXiv:1807.06209 [astro-ph.CO]}
  \BibitemShut {NoStop}%
\end{thebibliography}%

\end{document}